\documentclass{aa}  

\usepackage{graphicx}
\usepackage{txfonts}
\usepackage[switch]{lineno}
\usepackage[flushleft]{threeparttable}
\usepackage[normalem]{ulem}

\newcommand{\xWater}{x_{\rm H_2O}}
\newcommand{\Tdmx}{T_{\rm dmx}}
\newcommand{\Tad}{T_{\rm ad}}
\newcommand{\Tone}{T_{\rm 1\,\rm bar}}

\newcommand{\Zatm}{Z_{\rm atm}}

\usepackage{xcolor}

\newcommand\redsout{\bgroup\markoverwith{\textcolor{red}{\rule[0.5ex]{2pt}{0.4pt}}}\ULon}

\begin{document}

   \title{$\rm H_2-H_2O$ demixing in Uranus and Neptune: Adiabatic structure models}


   \author{M. Cano Amoros
          \inst{1}
          ,
          N. Nettelmann
          \inst{2}
          ,
          N. Tosi
          \inst{1}
          ,
          P. Baumeister
        \inst{3}
        ,
        H. Rauer
        \inst{1}
          }

   \institute{Institute of Planetary Research, German Aerospace Center (DLR),
Rutherfordstr. 2, Berlin, 12489, Germany\\
         \and
             Department of Astronomy and Astrophysics, University of California
(UCSC), High St 1156, Santa Cruz, 95064, CA, USA\\
         \and
         Department of Earth Sciences, Freie Universität Berlin, Malteserstr. 74-100, 12249 Berlin, Germany\\
}

   \date{Received: 6 September 2024 / Accepted: 25 October 2024}

 
  \abstract
    {Demixing properties of planetary major constituents influence the interior structure and evolution of planets. Comparing experimental and computational data on the miscibility of hydrogen and water to adiabatic profiles suggests phase separation between these two components occurs in the ice giants Uranus and Neptune.}
    {We aim to predict the atmospheric water abundance and transition pressure between the water-poor outer envelope and the water-rich deep interior in Uranus and Neptune.}
    {We construct seven H$_{2}$-H$_{2}$O phase diagrams from the available experimental and computational data. We compute interior adiabatic structure models and compare these to the phase diagrams to infer whether demixing is occurring.} 
    {We obtain a strong water depletion in the top layer due to rain-out of water and find upper limits on the atmospheric water mass fraction $\Zatm$ of 0.21 for Uranus and 0.16 for Neptune. The transition from the water-poor to the water-rich layer is sharp and occurs at pressures $P_Z$ between 4 and 11 GPa. Using these constraints on $\Zatm$ and $P_Z$, we find that the observed gravitational harmonics $J_2$ and $J_4$ can be reproduced if $P_Z\gtrsim 10$ GPa in Uranus and $\gtrsim 5$ GPa in Neptune, and if the deep interior has a high primordial water mass fraction of 0.8, unless rocks are also present. The agreement with $J_4$ is improved if rocks are confined deeper than $P_Z$, for instance below a rock cloud level at 2000 K (20--30 GPa). }
    {These findings confirm classical few-layer models and suggest that a layered structure may result from a combination of primordial mass accretion and subsequent phase separation. Reduced observational uncertainty in $J_4$ and its dynamic contribution, atmospheric water abundance measurements from an Orbiter with a Probe mission to Uranus (UOP) or Neptune, and better understanding of the mixing behaviour of constituents are needed to constrain the interiors of ice giants.}

    \authorrunning{Cano Amoros et al.}
    \titlerunning{}
    \maketitle{}
%

\section{Introduction}

    To understand the origin of the solar system and how planets form, it is important to unveil the mysteries around the interiors of the outer planets Uranus and Neptune as they are giant reservoirs of the planet building blocks in the early outer solar system. Modelling their interior structure is one of the means in that direction. 
    Uranus and Neptune are termed ice giants because their deep interior densities are consistent with that of a compressed mix of the ice-forming volatiles water, methane, and ammonia, or even of water-only if hydrogen-rich gas is assumed to be present in their deep interiors as well. Oxygen (O) and carbon (C) are the next most abundant elements after hydrogen (H) and helium (He) in the protosolar disk gas. They condense to water or CO ice in the outer regions of protosolar disks \citep{Mousis24}. If exposed to accreted H-rich gas, C and O react to form the ``ices'' water and methane. Therefore, the chemistry and phase diagrams of H-, O-, and C-mixtures are expected to be important for understanding ice giant interiors. 

    Interior structure models are constrained by observations, foremost by planetary mass, radius, atmospheric temperature, gravitational harmonics, and rotation rate. For the ice giants, the accuracy of these quantities is still limited compared to the gas giants Jupiter and Saturn. With Juno and the Cassini Grand Finale Tour, the gas giants had their dedicated orbiter missions to measure the gravity field \citep{Bolton17,Iess19}. At Jupiter, the Galileo entry probe measured atmospheric composition. In contrast, the ice giants have only been visited by a single Voyager 2 flyby each. 
    
    Currently, Uranus and Neptune models point towards a H/He-rich atmosphere and a deeper interior enriched in heavy elements \citep{Helled_2020}. This outcome of interior models is robust and independent of the approach used to infer the interior mass distribution. Two main approaches can be distinguished. One approach assumes an adiabatic interior, consisting of distinct layers, usually a core composed of rocks, an ice-rich layer surrounding the core that primarily consists of O and H, and a top layer composed primarily of H and He \citep{HMacF80,nett2013}. The ice-rich layer may also contain the rock-forming elements Si, Mg, Fe or other volatiles such as C, N, and S if they are miscible \citep{Vazan22}. Physical equations of state (EoS) are used to describe the behaviour of these elements or their compounds at relevant pressure and temperature conditions. Another approach is to adopt empirical structure models without making a priori assumptions on the composition, using density profiles described by high-order polynomials \citep{helled11, mov2020}, polytropes \citep{neu22, neu24} or random monotonic functions \citep{podolak22}. The latter method facilitates the possibility of considering more complex structures, including compositional gradients and non-adiabatic interiors \citep{neu24,malamud24}. However, even if the profiles match observations, relating these profiles to physical EoSs remains challenging. 
    
    Regardless of the method used, several open questions remain that require physical principles to confirm, exclude, and justify possible density profiles, for example, whether layers undergo gradual transitions, whether the $P-T$ profiles are adiabatic implying convection, or even whether the ice giants are water-rich at all or perhaps rock-dominated instead with H-gas to reduce the density where needed \citep{HelledFortney20}. At any rate, all approaches consistently predict the outer parts of the two ice giants to have a lower metallicity ($Z$) than the deeper interior, which is rich in heavy elements. Explaining such a structure by physical principles is important for understanding the formation and evolution of the planets. 

\begin{figure}[ht]
   \centering
   \includegraphics[width=0.5\textwidth]{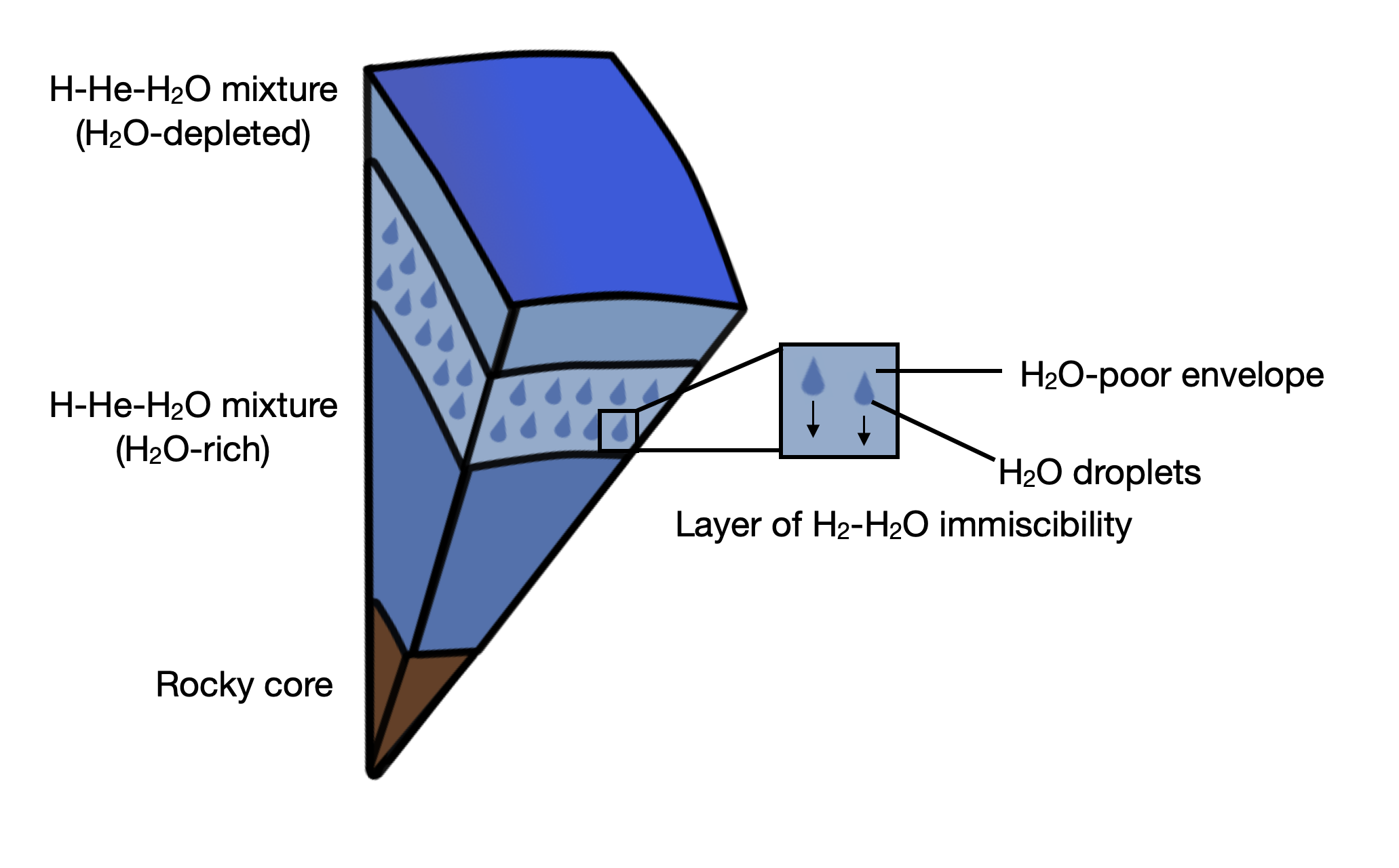}
   \caption{Illustration of the interior of an ice-giant-like planet with a H$_{2}$-H$_{2}$O immiscibility layer and water-rich droplets sinking and enriching the deeper envelope while leaving behind a water-depleted, hydrogen-rich top layer. This illustration also applies to He-rain in a gas giant planet, as shown in \cite{militzer2016}.}
      \label{fig:rain-out}
   \end{figure}

    \citet{Bailey_2021} provide a physical explanation for the traditional three-layer structure. They suggest that immiscibility (or demixing) between molecular hydrogen and water leads to a sharp compositional transition (Figure \ref{fig:rain-out}). To fit the gravitational harmonics, they find that Neptune requires a water abundance at least an order of magnitude higher in the top H-dominated envelope compared to Uranus. The authors explain this discrepancy by hypothesizing that Neptune may be in an earlier phase of H$_2$-H$_2$O demixing while Uranus would already be fully differentiated. Consequently, Neptune could be experiencing gravitational energy release of sufficient magnitude to account for its high heat flow today, thereby offering a plausible explanation for the observed dichotomy in the luminosities between the ice giants \citep{PEARL1990, pearl&conrath}. 
  
    Demixing is an important process also on other planets, for instance in Jupiter, where the separation of helium from metallic hydrogen at Mbar pressures and subsequent He-rain under the influence of gravity can explain the observed atmospheric helium and neon depletion \citep{zahn1998,wilson&mil2010}.
    However, experimental data on H-He demixing at Mbar pressures are sparse, and gas giant models rely on theoretical H-He phase diagrams with large uncertainties in the demixing temperature ($T_\text{dmx})$ of the order of 1000 K. 

    The mechanism proposed by \citet{Bailey_2021} for H$_2$-H$_2$O demixing in the ice giants is based on experimental data of the immiscibility of H$_2$-H$_2$O mixtures up to 3 GPa \citep{Seward1981,bali13}. Based on these data, \citet{Bailey_2021} constructed a H$_2$-H$_2$O phase diagram and found that their ice giants' adiabats were warmer than the phase boundary from the experimental data. Only upon linear extrapolation of this phase boundary beyond 3 GPa could \citet{Bailey_2021} find an intersection between the adiabats and the demixing curve, with consequent demixing in the interiors of Uranus and Neptune.

    Subsequent to \citet{bali13}, \citet{Vlasov23} conducted experiments on H$_2$-H$_2$O miscibility up to 4.5 GPa, and \citet{berg24} used DFT-MD simulations to obtain the H$_2$-H$_2$O phase diagram. Here, we use these new results as input to follow up on the proposal by \citet{Bailey_2021} that H$_2$-H$_2$O demixing may shape the structure of the ice giants that is suggested by the gravity data. 

    We construct phase diagrams constrained by the new data and various extrapolations thereof. By computing the equilibrium water abundance on the phase boundary we predict the atmospheric water abundance $Z_\text{atm}$ at the bottom of the water cloud deck. We determine the water-poor/water-rich transition pressure ($P_Z$) in the planet and construct interior structure models for Uranus and Neptune constrained by $Z_\text{atm}$ and $P_Z$ to compare with the observed gravity data. We find that rain-out can happen in the interiors of Uranus and Neptune and our inferred outer envelope water abundance and transition pressure leads to interior structures that can match the gravity harmonics.

    The remainder of the paper is organised as follows. Section 2 describes our construction of H$_2$-H$_2$O phase diagrams from the experimental and theoretical data.
    In Section 3 we apply these diagrams to predict $\Zatm$ and $P_Z$. 
    Section 4 presents Uranus and Neptune structure models constrained by $\Zatm$ and $P_Z$ for comparison to their observed gravity field.
    In Section 5 we discuss our results and finish with a summary of the conclusions in Section 6. In the Appendix, we present the systematic behaviour of the model gravitational harmonics upon variation of free parameters.

\section{H$_2$-H$_2$O phase separation under ice giant $P-T$ conditions}\label{methods}

\subsection{Ice giant adiabats}\label{subsec:adiabats}

    We assume adiabatic $P-T$ profiles for an envelope composed of hydrogen, helium and water, which will become divided into an inner and an outer part due to H$_2$-H$_2$O demixing.
    These profiles are obtained starting at a pressure level of 1 bar and a corresponding surface temperature $\Tone =T(P=1\;\text{bar})$. For hydrogen we use the effective H EoS and for helium the pure He EoS by \cite{cd21}, which combines semi-analytical EoS models with ab-initio molecular dynamics. In comparison to \cite{chabrier19}, these updated EoSs account for interactions between H and He. We add H$_2$O to the mixture using the AQUA EoS from \cite{aqua2020}, which is a combination of various EoSs, and covers a wide pressure and temperature range (0.1 Pa to 400 TPa and 100 to $10^5$ K). 

    The three tabulated EoSs are mixed using the additive volume rule. Following equations (9)-(11) from \cite{chabrier19}, for mass abundances $X$, $Y$, $Z$ of H, He, and H$_2$O respectively, we calculate the ideal mixing entropy to obtain the specific entropy of the mixture. Based on the computed mixture, we calculate the adiabat by interpolating the entropy $S(T,P,X_i)$ for a mixture $X_i$ starting at a given value of $\Tone$. As the tables do not cover temperatures lower than 100 K, we employ ideal gas EoSs for H, He and H$_2$O for $\Tone < 100$ K as done by \citet{scheibe19}. Compared to the physical EoSs of \citet{cd21}, the ideal gas approximation is appropriate at pressures of $\sim$ 1-100 bar. We set the switch to non-ideal EoS at 10 bar since this pressure level was found to produce smooth $P-T$ profiles and the temperatures along the adiabats have passed the 100 K threshold. For the iron-rock core, we use the pressure-density relation of \cite{hubbard1989} for an Earth-like bulk abundance of 38\% SiO$_2$, 25\% MgO, 25\% FeS and 12\% FeO by mass.

\begin{figure}[ht!]
   \centering
   \includegraphics[width=0.5\textwidth]{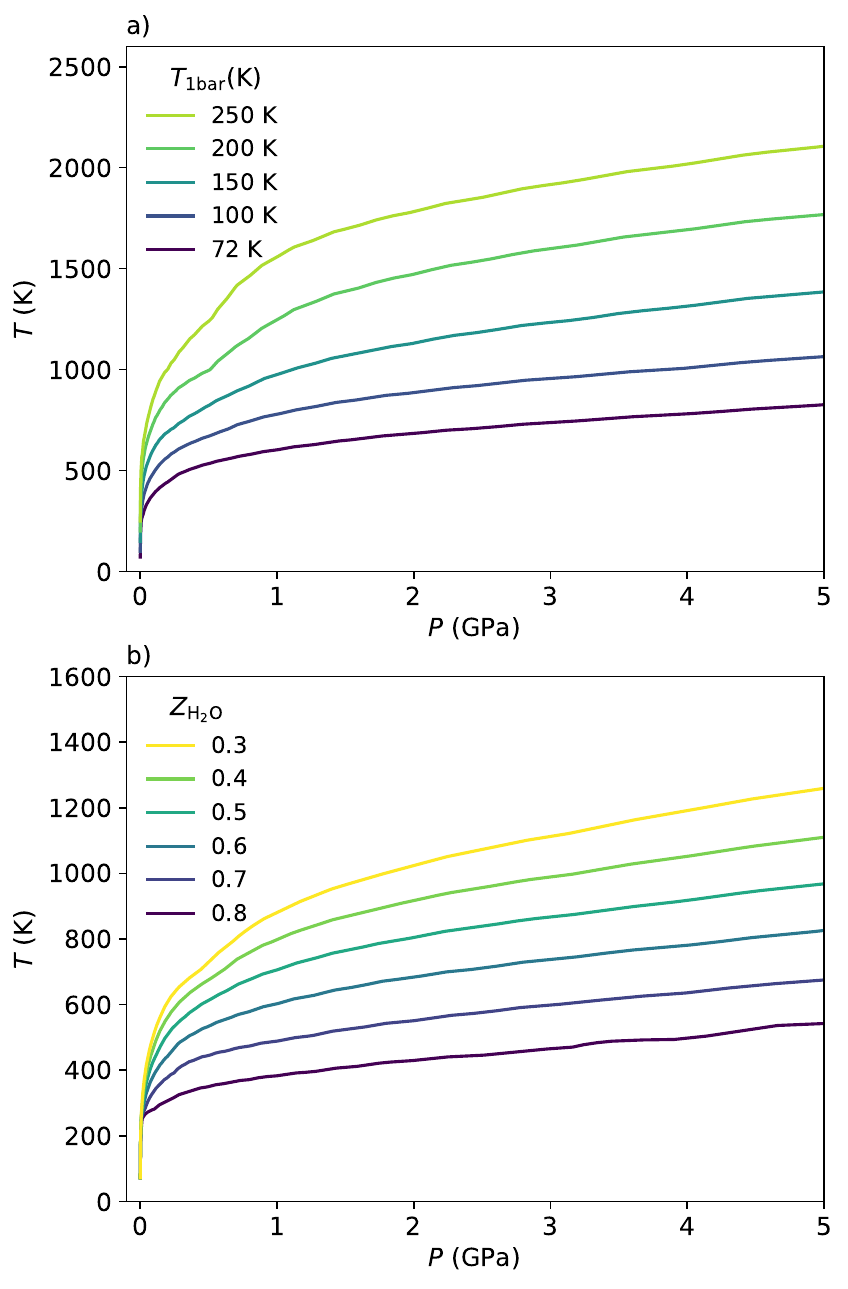}
   \caption{Adiabatic temperature profiles for (a) different surface temperatures $\Tone$ and fixed $Z_{\rm H_2O} = 0.6$ and (b) different $Z_{\rm H_2O}$ with $\Tone$ fixed at Neptune's value of 72 K. For each case the He/H ratio was kept protosolar. }
      \label{fig:ads}
   \end{figure}

    Figure \ref{fig:ads} shows examples of adiabatic profiles for different values of $\Tone$ between 72 K and 250 K and a fixed water mass fraction $Z_{\rm H_2O} = 0.6$ (Figure \ref{fig:ads}a), or for Neptune's surface temperature of $\Tone=72$ K but varying water mass fraction between 0.3 and 0.8 (Figure \ref{fig:ads}b). Figure \ref{fig:ads} shows that higher water abundances lead to colder adiabats and that the change in the gradient is strongest at pressures below $\sim 1$ GPa. The behaviour of the adiabats with increasing water abundance can be explained by the inverse relation between the adiabatic gradient and the specific heat capacity, which is larger for molecules with higher degrees of freedom.

\subsection{Construction of H$_2$-H$_2$O phase diagrams}\label{sec:phasediagram}

\begin{figure}[]
   \centering
   \includegraphics[width=0.5\textwidth]{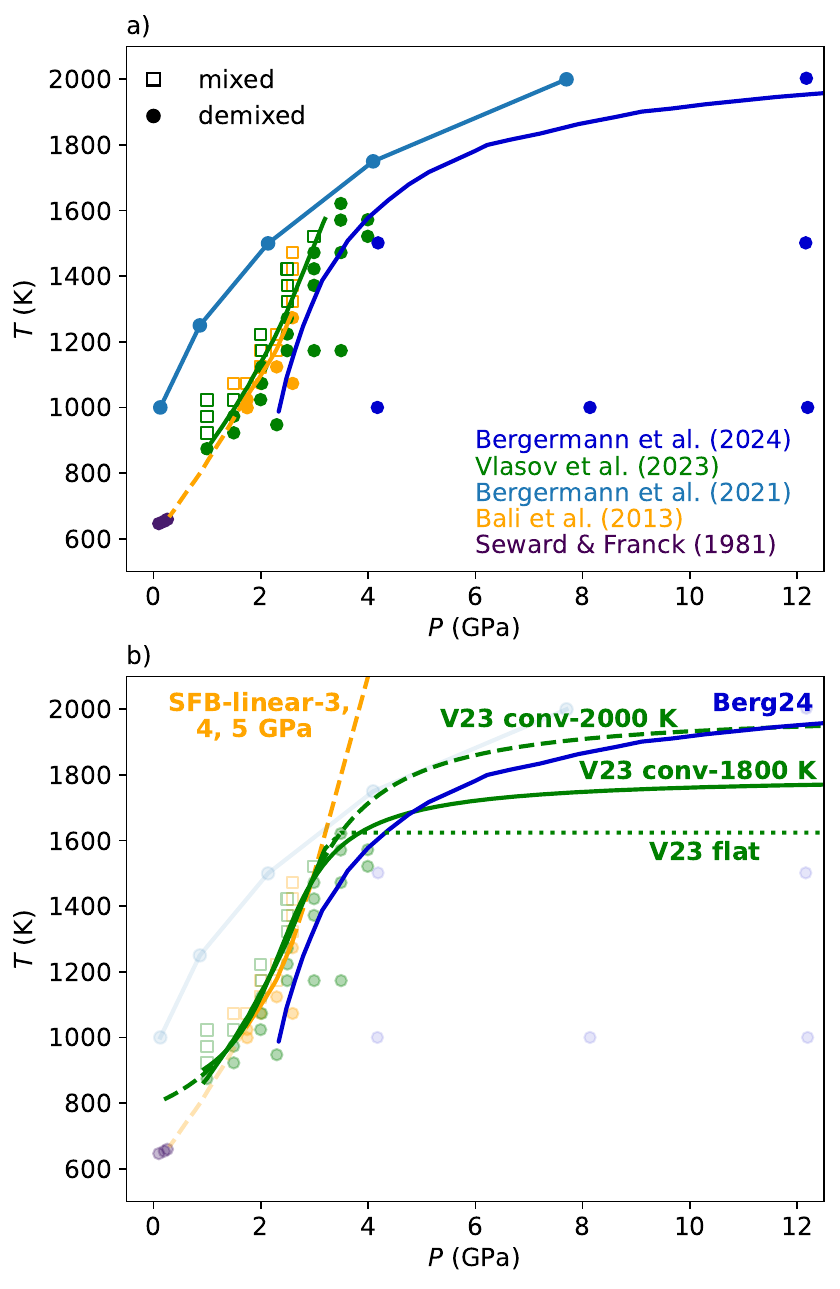}
   \caption{Experimental data and theoretical predictions on H$_2$-H$_2$O miscibility. a) Experimental data for 1:1 H$_2$-H$_2$O miscibility from \cite{Seward1981} (purple), \cite{bali13} (yellow), \cite{Vlasov23} (green) and the computational predictions of \cite{berg21} (light blue curve) and \citet{berg24} (blue curve). Filled symbols correspond to the coexistence of two phases while empty squares to complete mixing of H$_2$ and H$_2$O. The solid lines indicate the fit of the corresponding data points. b) Extrapolated critical curves based on the data shown in panel a) and used to construct the phase diagrams (see text for details). The linear extrapolations of the data by \cite{bali13} beyond 3 GPa to 4 and 5 GPa are shown by the dashed yellow line. The three different extensions beyond 3.5 GPa for the \cite{Vlasov23} data (flat, V23 conv-1800 K, and V23 conv-2000 K) are shown by green dotted, solid and dashed lines, respectively. Finally, the blue curve is from \cite{berg24}.}
      \label{fig:crit_curve}
   \end{figure}

    To construct a H$_2$-H$_2$O phase diagram, we rely on experimental immiscibility data from \citet{Seward1981}, \citet{bali13}, and \citet{Vlasov23}, as well as on theoretical predictions by \citet{berg21} and \citet{berg24}. This collection of data is shown in Figure \ref{fig:crit_curve}a.

    \citet{Seward1981} (hereafter SF81) study the immiscibility of H$_2$-H$_2$O up to 2500 bar (0.25 GPa) and 440 °C. They conduct isochoric experiments for H$_2$-H$_2$O mixtures ranging from hydrogen-poor (0.5 mol\%) to hydrogen-rich (90 mol\%) using a pressure vessel. The critical, or demixing temperature ($\Tdmx$) is the maximum temperature at which a mixture can co-exist as two distinct phases, with higher temperatures leading to complete mixing. Figure \ref{fig:crit_curve}a shows in purple the critical $\Tdmx$-points of \cite{Seward1981} at 0.1, 0.2 and 0.25 GPa for a 1:1 H$_2$-H$_2$O bulk mixture. 

    \citet{bali13} (hereafter B13) conducted experiments at higher pressures than SF81 using the synthetic fluid inclusion method. By trapping fluids in silicate minerals they study the immiscibility of a 1:1 H$_2$-H$_2$O mixture up to $3$ GPa. They infer the location of the critical curve for the mixture by visually inspecting the inclusion type and confirming its composition with Raman spectroscopy. Inclusions containing both a H$_2$- and a H$_2$O-rich phase were interpreted as cases of full miscibility, while compositionally different inclusions were interpreted as cases of immiscibility. Their experimental data are shown in Figure \ref{fig:crit_curve}a as yellow dots and their fit by the solid yellow line.

    We use the above-mentioned experimental data to construct the first three of our demixing curves $\Tdmx(P)$ for a 1:1 mixture. Figure \ref{fig:crit_curve}a shows with a dashed yellow line that the critical temperatures of SF81 connect well to the data of B13, as also shown by \cite{Bailey_2021}. The curve is nearly linear, with a slight change in slope between 2.5 and 3 GPa. As in \citet{Bailey_2021}, to reach conditions at higher pressures relevant to the ice giants' interiors, we extrapolate linearly this 1:1 critical curve toward higher pressures beyond 3 GPa. We cut off the critical curves at 3 GPa (no extrapolation), 4 GPa, and 5 GPa, and label the three corresponding phase boundaries and subsequent phase diagrams as ``SFB-linear-3 GPa'',  etc. The cutoff implies that we assume miscibility at higher pressures beyond the respective cutoff pressure. These extrapolated critical curves are shown in Figure \ref{fig:crit_curve}b. They serve to compare our results with those of \cite{Bailey_2021} and to explore the response of the models in the absence of experimental data beyond 3 GPa, i.e. prior to the arrival of the Vlasov data.

    More recently, \cite{Vlasov23} (hereafter V23) obtained new experimental data for the critical curve of the H$_2$-H$_2$O system up to 3.5 GPa and 1400 °C following the same method as \cite{bali13}. For 4 GPa, they provide only a lower limit to the critical temperature.  At $\sim$3.5 GPa, their data indicate a flattening of the critical curve. We use this additional information to construct three more, improved critical curves $\Tdmx (P)$ beyond 4 GPa. First, we take a ``flat'' case, for which we extend the curve from the last critical temperature measurement at $\sim$3.5 GPa horizontally toward higher pressures. This yields the lowest possible $\Tdmx(P)$ 1:1 curve. Additionally, we aim to approximate the change of slope indicated by the V23 data and the theoretical prediction of \cite{berg21} (see next paragraph). To do so, we fit the experimental data of B13 and V23 by an arctan function that approaches either 1800 K at high pressures (case ``V23 conv-1800 K'') or 2000 K (case ``V23 conv-2000 K''). In contrast to the linear extrapolations, these three critical curves are consistent with the theoretical predictions by \citet{berg21}.

    \citet{berg21} perform Gibbs ensemble Monte Carlo simulations up to 8 GPa and 2000 K with analytical pair-potentials for the interaction between the molecular species. This approach neglects the variable electronic structure of the molecules, which they suggest is what leads to the difference with the data of \cite{bali13}. In contrast, in \cite{berg24} (hereafter Berg24), DFT-MD simulations are employed which do include the electronic structure within the density-functional theory for the electronic subsystem. This apparently leads to a downward shift in $\Tdmx$ and a good agreement with the experimental data at a few GPa. In both theoretical cases, the critical curves bend over toward higher pressures and do not exceed 2000 K. Our V23 conv-1800 K and V23 conv-2000 K cases precede the Berg24 results and, as Figure \ref{fig:crit_curve}b shows, bracket the Berg24 curve at $P>5$ GPa, while they follow the experimental points between 2 and 3.5 GPa. Finally, we adopt the critical curve of Berg24. This critical curve is slightly off compared to the experimental data up to 3 GPa. However, as will be shown later, this low-pressure region does not influence our phase separation results.  

\begin{figure}[]
   \centering
   \includegraphics[width=0.5\textwidth]{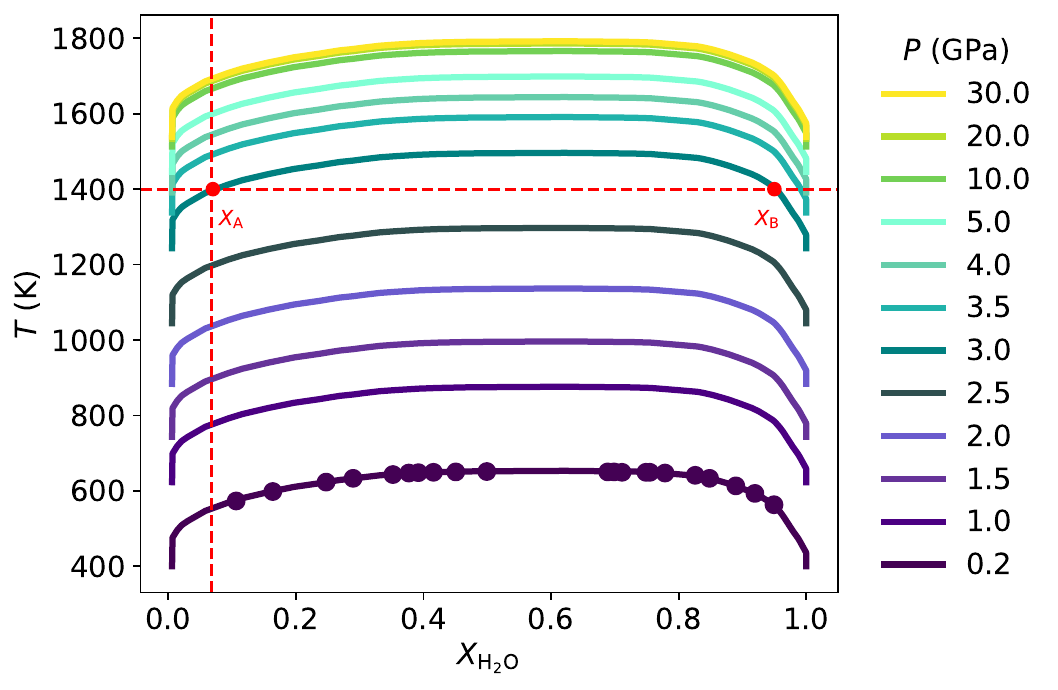}
   \caption{Example of a phase diagram $\Tdmx(P,\xWater)$ based on the shape of the 0.2 GPa isobar according to the experimental data (purple points) of \cite{Seward1981}. Higher isobars are obtained by shifting the 0.2 GPa curve vertically according to the given critical curve $\Tdmx(P,0.5)$. Here the V23 conv-1800 K case is shown. $x_A$ and $x_B$ show the coexisting particle fractions at 3 GPa and 1400 K. The water-poor equilibrium abundance $x_A$ is used to compute the atmospheric abundance.}\label{fig:phase}
   \end{figure}

    In summary, we have seven critical curves at our disposal: SFB-linear-3, -4, -5 GPa, V23 flat, V23 conv-1800 K, V23 conv-2000 K and Berg24. From these curves, we construct seven H$_2$-H$_2$O phase diagrams. The shape of the phase diagram is based on results from \cite{Seward1981}. They find that the isobaric curve at 0.2 and 0.25 GPa becomes nearly symmetric around the 1:1 H$_2$-H$_2$O concentration, where $\Tdmx$ attains a maximum. This symmetric behaviour is observed at 0.2 and 0.25 GPa. The experimental data for the 0.2 GPa isobar is shown in purple dots in Figure \ref{fig:phase}. The way we construct phase diagrams from the sparse experimental or theoretical data described above is by assuming that the shape of the isobaric demixing curve observed at 0.2 GPa holds for all higher pressures considered. Therefore, given a point $\Tdmx(P, \xWater)$, for any higher pressure $P$ and any water concentration $\xWater$, we can readily draw the respective isobaric demixing curve $\Tdmx(P, \xWater)$ by shifting the baseline curve for 0.2 GPa upwards with temperature. An example of a resulting phase diagram is illustrated in Figure \ref{fig:phase} for the critical curve V23 conv-1800 K. In the same manner, we construct phase diagrams for the rest of the critical curves. 
    In the Berg24 case, a phase diagram consistent with the experimental data up to 3 GPa and the Berg24 curve beyond 4 GPa could have also been easily constructed via interpolation. However, we did not do this since this lower-pressure region is above the region where phase separation occurs, as will be shown in Section \ref{sec:Ptang}.

\subsection{Equilibrium water abundance}\label{sec:methZatm}

    For a given adiabat $\Tad(P,Z)$ of composition $Z$ and protosolar ratio $X$/$Y$, constrained by $\Tone$, and a phase diagram $\Tdmx(P,Z)$, phase separation occurs when $T_\text{ad}(P,Z) < T_\text{dmx}(P,Z)$. Water-rich droplets rain out, and $Z$ decreases until $T_\text{ad}(P,Z_A) \geq T_\text{dmx}(P,Z_A)$ at all pressures, where $Z_A$ is the equilibrium mass fraction on the water poor-side of the phase diagram (see Figure \ref{fig:phase}). Equality holds where the adiabat is tangential to the demixing curve, which typically occurs at a single pressure point but can also occur over an extended pressure region. Note that we use the symbol ``$Z$'' for mass fraction and ``$x$'' for particle fraction. Since in this work we only use water for the heavy elements, here $Z = Z_{\rm H_2O}$ and $x = x_{\rm H_2O}$. The conversion between $Z_{\rm H_2O}$ and $\xWater$ is done using Equation \eqref{eq:xZconversion} below. $\Tdmx(P,\xWater)$ is what is read from the phase diagram (Figure \ref{fig:phase}) and converted into $T_\text{dmx}(P,Z_{\rm H_2O})$. We then compare the adiabatic profile $\Tad(P,Y,Z)$ and the demixing curve $\Tdmx(P,Z)$ to check whether demixing is occurring and what the resulting equilibrium water abundance due to rain out is. In Figure \ref{fig:phase} the equilibrium composition $x_A$ on the water-poor side of the phase boundary is shown for an example pressure of 3 GPa and temperature of 1400 K. To determine $x_A$ (or $Z_A$), we implemented a new procedure in our interior structure code TATOOINE (see Section 3.1), which is similar to the procedure adopted by \citet{nett15,nett24} to treat H-He demixing. 
  
\begin{figure}[]
   \centering
   \includegraphics[width=0.5\textwidth]{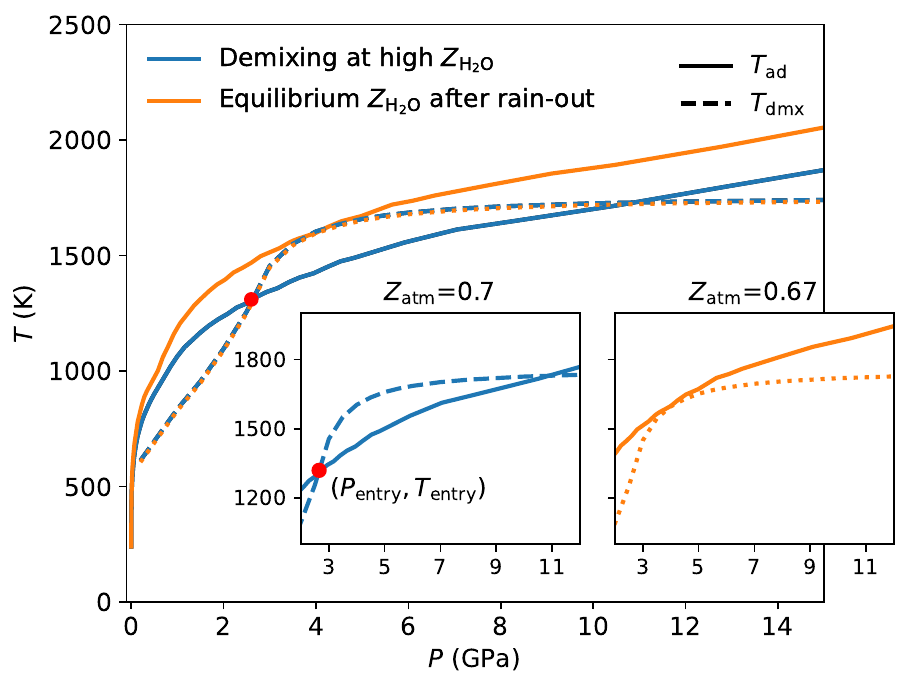}
   \caption{Determination of the water equilibrium abundance. Solid lines and dashed lines denote adiabats and demixing curves respectively, the latter obtained from the V23 conv-1800 K phase diagram shown in Figure \ref{fig:phase}. We start with an overlap region where $T_\text{ad} < \Tdmx$ and an initial water abundance $Z=0.7$ throughout the entire planet. Then $Z$ is decreased until the overlap region disappears and $T_\text{ad}\geq\Tdmx $ at all pressures. Here the equilibrium abundance found is 0.67. The orange adiabat is the profile with a 0.67 outer envelope abundance. The obtained abundance is high in this example because here $\Tone=240$ K implying a rather warm adiabat in a young planet. The demixing region for the case with $Z_\mathrm{atm}=0.7$ is much larger compared to that typically encountered in our simulations. This choice was made for visual clarity. See text for more details.} \label{fig:convergence}
   \end{figure}

    In the case of H-He phase separation, both the adiabat and the demixing curve become colder with decreasing He-abundance but at a different rate, which is how in that case an equilibrium He-abundance can be found.  Here, upon water rain-out, the adiabat becomes warmer while the demixing curve changes little. This behaviour is illustrated in Figure \ref{fig:convergence}.
    For a high water abundance $Z = 0.7$ (blue curves), a wide overlap region where $\Tad < \Tdmx$ exists. From the pressure at the entry of the overlap region ($P_\text{entry}$), we obtain a new isobar of our phase diagram by interpolation.  From the corresponding temperature ($T_\text{entry}$), we obtain the two co-existing abundances, $x_{A}$ and $x_{B}$ in particle fraction (as shown in Figure \ref{fig:phase}), or $Z_{A}$ and $Z_{B}$ in mass fraction as follows
\begin{equation}\label{eq:xZconversion}
Z_{\rm H_2O} = \frac{9\:\xWater}{1 + 8\:\xWater} \; ,
\end{equation}
which is valid only in the absence of He. In the presence of He with a protosolar mass fraction $Y$, we have
\begin{equation}\label{eq:Z}
Z_{\rm H_2O} = \frac{m_{\rm H_2O}\:{\rm O/H} }{ m_{\rm H_2O}\: {\rm O/H} 
+ m_{\rm He}\:{\rm He/H} 
+ m_{\rm H} \: (1-2{\rm O/H})}\;,
\end{equation}
with
\begin{eqnarray}\label{eq:He_H}
{\rm He/H} &=& \frac{\rm He/H_{free} }{\rm 1 + 2O/H_{free}}\;, \\
{\rm He/H_{free}} &=& \frac{Y}{1-Y} \: \frac{ m_{\rm H}}{m_{\rm He} }\;,\\
\rm O/H_{free} &=& \rm \frac{O/H }{ 1-2O/H}\;,
\end{eqnarray}
where H$_\text{free}$ refers to H atoms that are free or bound in H$_2$, while H in X/H refers to all H atoms. We take $Z_{A}$ as the new water abundance of the adiabat.

    With a lower $Z$ in the outer envelope, the new adiabat is slightly warmer. We repeat the whole procedure iteratively until $\Tad$ and $\Tdmx$ intersect only once, as shown in Figure \ref{fig:convergence} by the orange curves. We have gone from having an overlap region between $\Tad$ and $\Tdmx$ with $Z=0.7$ to the point where $\Tad \approx \Tdmx$ when water has rained out and the water abundance has decreased to the equilibrium value $Z=0.67$. We point out that the adiabats in Figure \ref{fig:convergence} do not show two consecutive steps in our procedure. Since we start our procedure at the first instance in temperature where demixing can occur, we never have such a large demixing region as shown in the 0.7 case but instead, are always close to equilibrium. 
    Therefore, the exaggerated temperature increase that can be seen here does not actually occur. 
    
    Furthermore, one can see from the dashed lines in Figure \ref{fig:convergence} that the change in the demixing curve with $Z$ is tiny. This can be explained by the rather flat behaviour of the isobars in the phase diagram throughout most of the compositional range (see Figure \ref{fig:phase}).
    We take this equilibrium abundance as the atmospheric water abundance ($\Zatm$). In this way, we assume that convection will cause mixing of all material above the rain-region and thus the top envelope will adopt this abundance because the convective overturn happens faster than the rain-out in this region.

\section{Models based on H$_2$-H$_2$O phase diagrams}

    We compute four kinds of structure models: (i) models where the atmospheric water abundance $\Zatm$ and the transition pressure $P_Z$ are constrained by the H$_2$-H$_2$O phase diagrams, (ii) models where, in addition, the deep interior water abundance is constrained by the gravitational harmonic $J_2$, (iii) models also constrained by $J_2$ but with water and rocks in the deep interior above the core, and (iv) models that are constrained by $J_2$ only. In Section 3.1, we describe case (i) models and their results in Section 3.2. Models of cases (ii-iv) are described in Section 4.1. 

\subsection{Structure models constrained by water rain-out}


    We follow the traditional approach of modelling the planet's interior assuming a layered structure (\citealp[e.g.][]{nett2013}). We use the 1D interior structure code TATOOINE \citep{Baumeister_2020,MacKenzie_2023,Baumeister_2023} to construct a planet consisting of an isothermal core and two adiabatic envelopes on top. The envelopes are composed of H and He in a protosolar ratio, and H$_2$O. The model inputs are the planet mass, the mass fractions for each layer, and the equations of state (mentioned in Section 2.1) describing the properties of the relevant elements at high pressure and temperature conditions. TATOOINE integrates numerically the equations of mass continuity \eqref{mass-con} and hydrostatic equilibrium \eqref{hydro}, coupled with the equation of state \eqref{eos}: 
\begin{align}
& \frac{\mathrm{d} m(r)}{\mathrm{d} r} = 4\pi r^{2}\rho(r)\;,  \label{mass-con}\\
& \frac{\mathrm{d} P(r)}{\mathrm{d} r} = -\frac{Gm(r)\rho (r)}{r^{2}}\;, \label{hydro}\\
& P(r) = f(\rho (r), T(r), c(r)) \;,\label{eos}
\end{align}
where $r$ is the radius, $\textit{m}$ is the mass, $\rho$ is the density, $P$ is the pressure, $T$ is the temperature, and $c$ is the composition. The above equations are integrated iteratively from the top downward until the final solution is obtained when the mass at the planet's centre reaches approximately zero. 
In these models, we fix the rock core mass at 2$M_E$.


    For any temperature at the top of the atmosphere (here 1 bar pressure level) $\Tone$, we find the transition pressure, $P_Z$, between the water-poor and water-rich envelope. Given $P_Z$, for each $\Tone$, we determine $\Zatm$ and the mass from the center of the planet up to $P_Z$, $m_Z:=m(P_Z)$ from the planet structure model. Envelope 1 extends between $M_p$ and ${m}(P_Z)$, and envelope 2 begins at ${m}(P_Z)$. The first structure model is homogeneous, i.e. $Z = Z_1 = Z_2$ where $Z_1$ and $Z_2$ denote the water abundances of envelopes 1 and 2 respectively. Once we find at what surface temperature demixing can start, all parameters are known and the model where phase separation starts can be computed. For a colder adiabat (i.e. a lower $\Tone$), $Z_1 < Z_2$ due to rain-out. The new $Z_1$ is determined from the phase diagram as described in Section 2.3. We compute the mass of water $\Delta m_{{\rm H}_2{\rm O}}$ that has rained out from envelope 1 to envelope 2. Given the previous amounts of water in the two envelopes and the change $\Delta m_{{\rm H}_2{\rm O}}$, we compute the new value $Z_2$ according to bulk water mass conservation. With these new abundances for the envelopes and $P_Z$, a new structure model is computed which yields the new $m_Z$. This is done repeatedly for various $\Tone$ values. We find that $m_Z$ decreases as the planet cools. This is the result of two opposing effects: Envelope 1 becomes less dense and warmer as it loses water. This tends to decrease $m_Z$. Second, $m_Z$  would increase as the planet becomes more compact upon cooling \citep{nett2013}. We find that the first effect dominates, meaning that already rained-out material is distributed back into the outer envelope.

    By repeating this procedure until the present-day $\Tone$ value is reached, these models provide the atmospheric water abundance and transition pressure for present Uranus and Neptune, for the seven different H$_2$-H$_2$O phase diagrams. However, the deep interior-$Z$ is not directly constrained by the H$_2$-H$_2$O phase diagram, only its change according to mass conservation for an assumed initial homogeneous-$Z$ is determined.

\subsection{Results}

\subsubsection{Atmospheric water abundance}

\begin{figure}[!tbp]
   \centering
   \includegraphics[width=0.5\textwidth]{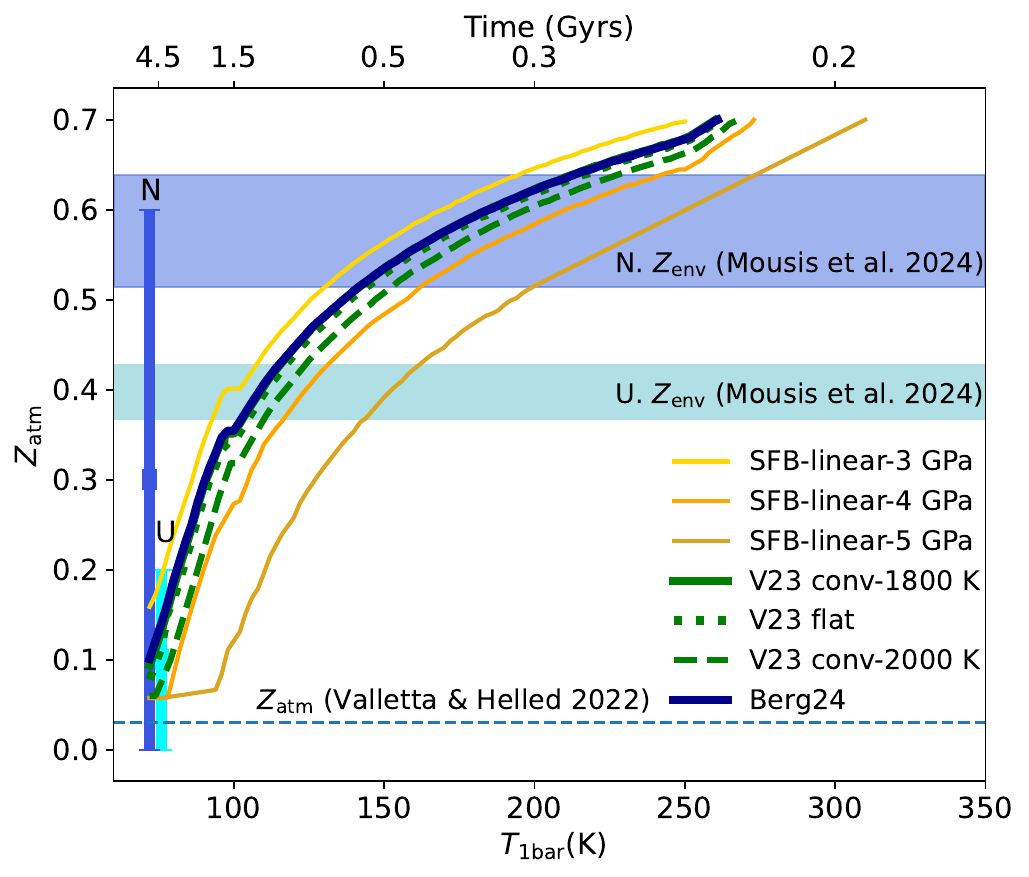}
   \caption{Atmospheric water abundance as a function of $\Tone$ for each of the seven phase diagrams. The V23 conv-1800 K curve is hidden by the Berg24 curve. The change of slope near 100 K is due to the switch to the ideal-gas EoS for $T<100$ K (see Sec. \ref{subsec:adiabats}). Vertical uncertainty ranges on the left show the outer envelope water abundance from the structure models of \citet{nett2013}. The top horizontal axis shows times corresponding to a model of Neptune's thermal evolution from \cite{nett2013}. The dashed horizontal line shows the metallicity of the outer envelope found by \citet{vallettahelled2022} using planet formation models. The horizontal shaded areas show the bulk oxygen enrichment predicted by \cite{Mousis24} for the envelopes of Uranus and Neptune ($Z_\text{env}$), which we converted to mass fractions.}
  \label{fig:zatm}
   \end{figure}

    We estimate the atmospheric water abundance $\Zatm$ for a range of 1 bar temperatures from the start of the demixing to the present-day temperatures, $\Tone$,  of Uranus and Neptune, which is the only parameter by which Uranus and Neptune are distinguished here. Figure \ref{fig:zatm} shows $\Zatm$ as a function of $\Tone$.  
    There are seven curves, one for each of the seven constructed phase diagrams. The vertical error bars show the range of water abundance in the outer envelope of Uranus and Neptune obtained from three-layer structure models of \citet{nett2013}.
  
    We start with a warm planet with a mixed interior of an arbitrary initial water abundance $Z = 0.7$. As the planet cools, the atmospheric water abundance decreases due to the rain-out of water. The colder the adiabat, the more water rains out. The demixing process depletes the outer envelope of water whilst enriching the deeper interior. For Uranus ($\Tone=76\:$K), we find a range for the atmospheric water abundance $\Zatm$ between 0.057 and 0.21. For Neptune ($\Tone=72$\:K), we obtain a range of $Z_\text{atm}$ between 0.057 and 0.16. For comparison, we show in Figure \ref{fig:zatm} the outer envelope heavy element metallicity from the formation models of \citet{vallettahelled2022}, which also consider only water. We also provide estimates of the bulk envelope oxygen enrichment from \cite{Mousis24} converted into mass fractions using Equations (2)-(5). These estimates put our arbitrary initial water abundance into context with regards to formation models.

    The time scale shown on the upper axis of Figure \ref{fig:zatm} is based on the evolution models of \cite{nett2013} and suggests that phase separation may have started early in the evolution of these planets when $\Tone$ temperatures were around 300 K. 

    Figure \ref{fig:zatm} shows the depletion of atmospheric water from the arbitrarily chosen starting abundance of 0.7. However, if the starting abundance were lower, as suggested by \citet{Mousis24}, demixing would start later on in the evolution of the planet, or earlier if the initial abundance were higher. This is due to the behaviour of the adiabat with water abundance as shown in Figure \ref{fig:ads}b. A higher initial water abundance will require a higher $\Tone$ for demixing to start. But in any case, the final atmospheric water abundance is independent of the starting initial abundance $Z$. This is because the outer envelope abundance required for the adiabat and demixing curve to intersect only once is always the same for each $\Tone$. Therefore, to set a constraint on the deep interior water abundance based on our $\Zatm$ predictions, interior structure models are required that are constrained by the observed gravitational harmonics.

\subsubsection{Transition pressure}\label{sec:Ptang}

    In H$_{2}$-H$_{2}$O demixing, the slope of the adiabat of the planet tends to be steeper than the slope of the demixing curve, and the latter changes little with $Z$. Therefore, both keep diverging for higher $Z$-values along the adiabat. While the temperature along the adiabat can be reduced through enhancement of $Z$, this is efficient only in the $<0.1$ GPa region (see  Figure \ref{fig:ads}b). Once pressures of the order of a few GPa are reached, the temperature along the adiabat is less affected by changes in $Z$. Therefore, no deeper equilibrium point exists. The rain region will be thin, and therefore characterized by a sharp water-poor/water-rich transition. This is different from the case of H-He demixing, where higher He-abundances lead to warmer demixing curves $\Tdmx(P, Y)$ as shown in \citet{nett15}. In that case, there can be a zone below the onset pressure of demixing where higher equilibrium He-abundances can be found at higher $P-T$ values along the adiabat, and thus the He-poor/He-rich transition becomes gradual \citep{nett15,Mankovich16,HowardHelledAA24}.

\begin{figure}[]
   \centering
   \includegraphics[width=0.5\textwidth]
  {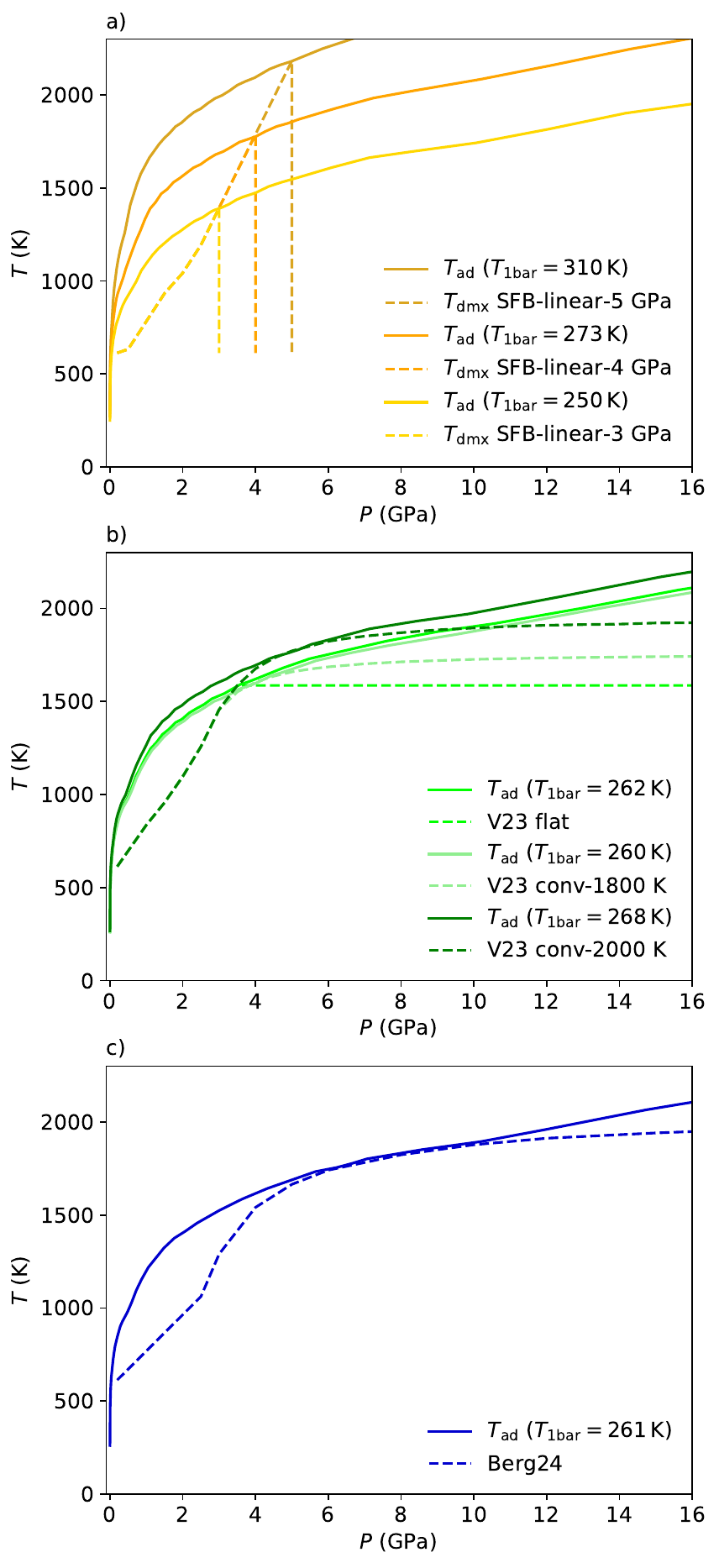}
   \caption{Adiabats (solid lines) and demixing curves (dashed) for the seven phase diagrams at the onset of demixing. All adiabats and demixing curves are for $Z = 0.7$. (a): SFB-linear-3, 4, 5 GPa, (b): V23-flat (lime), V23 conv-1800 K (light green), V23 conv-2000 K (dark green), (c): Berg24. }
   \label{fig:demix_start}
   \end{figure}

    We show the tangential behaviour for the seven phase boundaries in Figure \ref{fig:demix_start} and under the conditions of a young, warm planet, where demixing is just about to begin. 

    In the three SFB-linear cases, the intersection between the adiabat and the demixing curve occurs at the cutoff pressure due to the steep (linear) slope and abrupt drop of the demixing curves at higher pressures (Figure \ref{fig:demix_start}a). In these cases demixing starts at $\Tone\approx 250$ K for 3 GPa, 273 K for 4 GPa, and 310 K for 5 GPa for an initial water abundance of 0.7 that is homogeneous throughout the planet.

    With the three phase curves based on the V23 data, the planet adiabat for the equilibrium abundance and the demixing curve is tangential at pressures $P_Z\sim 4$ GPa. Upon cooling, $P_Z$ remains nearly constant. This is because although $Z$ decreases with $\Tone$ (Figure \ref{fig:zatm}), the cooling effect on the adiabat with $\Tone$ (Figure \ref{fig:ads}a) is compensated by the warming effect due to decreasing $Z$ (Figure \ref{fig:ads}b), and the demixing curve changes little with $Z$. Assuming 0.7 as the initial homogeneous water abundance, H$_2$-H$_2$O phase separation starts at $\Tone = 262$ K for the V23 flat case, at 260 K for V23 conv-1800 K, and at 268 K for V23 conv-2000 K (see Figure \ref{fig:demix_start}b).

    Only for the Berg24 phase diagram we observe a different behaviour. Here, the region where the adiabat corresponding to the equilibrium abundance is tangential to the phase boundary extends over a pressure range of $4-11$ GPa. Therefore, the rain-out region is wider, but homogeneous as well. 
    Any change in water abundance will occur at pressures deeper than $P_Z$. Because of this extended region, in the Berg24 case $P_Z$ refers to the end of the tangential region.
    Higher water abundances in the deep interior will not be on the phase boundary, and must therefore be due to other factors such as the formation process. Since there are no further equilibrium points at higher pressure, the water-poor/water-rich transition must be sharp if caused by H$_2$-H$_2$O phase separation. 
    For the Berg24 phase diagram, demixing starts at 261 K (Figure \ref{fig:demix_start}c). For lower initial water abundances, the onset of demixing would occur later, i.e. at lower $\Tone$ values.

\subsubsection{Deep-$Z$ level over time}

    We find that the increase in the deep water abundance in envelope 2 due to the rain-out from the outer layer is negligible. The deep interior water abundance remains essentially primordial. This is a consequence of the low mass of the outer envelope. In present Neptune for instance, the 4 GPa-level occurs at $0.975$ of Neptune's mass.

\section{Models constrained by gravitational harmonics}\label{sec:structure}

    In this section, we describe the already mentioned models where the deep interior abundance is constrained by the gravitational harmonic $J_2$ in addition to the constraints on $\Zatm$ and $P_Z$ from the phase diagrams. Case (ii) models assume water only for heavy elements in the deep envelope, and case (iii) models have both water and rocks in the deep interior above the core at a 0.5x solar ice-to-rock (I:R) ratio, except in the case of Neptune using the Berg24 data where 0.5 did not provide a solution and we therefore used I:R of 1x solar. Case (iv) models are constrained by $J_2$ only. Results for these models are presented in Section 4.4. In all of these cases, the rock core mass is a free parameter that is used to adjust $M_p(R_p)$ to the values of Uranus and Neptune.

\subsection{Case (ii) and case (iii) models }

    These models are constrained by $\Zatm$ and $P_Z$ from H$_2$-H$_2$O phase separation as explained in the previous section, and additionally by the observed gravitational harmonic $J_2$. The composition of the deep interior must be determined in order to satisfy the latter constraint. To this end,  we introduce two model series, one for an ice-rich composition, case (ii), and another for an ice-rock composition, case (iii).

    In case (ii) we confine rocks to the core and assume that water is the only heavy component present in the deep envelope, i.e. $Z_2 = Z_{\rm H_2O}$. We vary $Z_2$ to fit $J_2$ and then check how well models that satisfy this constraint can also match the observed gravitational harmonic $J_4$. The seven constrained models of this class are marked by star symbols in Figure \ref{fig:Js}. 

    In case (iii) we allow rocks to be additionally present in envelope 2, which substantially reduces the resulting I:R ratio of the models. According to the condensation curves of MgSiO$_3$ and MgSiO$_4$, a condensation temperature of 2000 K occurs at 0.1 GPa and changes only slowly with pressure \citep{VissherMoses11}. Since condensation temperatures generally rise with pressure, extrapolation to higher pressures suggests that 2000 K is a lower limit to the condensation temperatures in giant planets. We thus use 2000 K as an approximate level of silicate clouds and consider rocks only below this temperature. The adiabats of Uranus and Neptune reach 2000 K at pressures of about 8--15 GPa. In this series of models, we fix the I:R ratio at 0.5x (or 1x) solar for a solar I:R of 2.7.

    To compute the gravitational harmonics we use the MOGROP code \citep{nett2012} based on the theory of figures up to the 4th order \citep{zharkov78}. We checked that using the theory up to the 7th order does not influence the inferred $Z_2$ values to any significant digits. For consistency with TATOOINE, which employs the AQUA water EoS \citep{aqua2020}, we use TATOOINE's adiabats up to $P_Z$ as input to the MOGROP code. Beyond $P_Z$, we use the H$_2$O-REOS for water and H/He-REOS.3 for H and He \citep{Becker14}.

\subsection{Case (iv) models}

    In these models, abundances of envelopes 1 and 2 above the core ($Z_1$, $Z_2$), I:R ratio, and $P_Z$ are all free parameters. These models are only constrained by planetary mass ($M_p$), radius ($R_p$), rotation rate, and surface temperature ($\Tone$). In models with rotational flattening, $R_p$ means that the models fit the literature value for the equatorial radius ($R_{\rm eq}$) at the 1 bar reference level, which is 25559$\pm$4 km for Uranus and 24764$\pm$15 km for Neptune \citep{lindal1987,lindal1992}.
    Due to their flexibility, these models are well suited to study how the gravitational harmonics vary in response to variations of single parameters. We present such a parameter study in Figure \ref{fig:freeparametersj2n} in the Appendix.

\subsection{Observed gravitational harmonics}

    For Uranus and Neptune, the lowest order harmonics $J_2$ and $J_4$ have been observed by the Voyager 2 spacecraft. Long-term monitoring of the orbital positions of their natural satellites allowed those values to be substantially refined. Here, we use the observed values of $J_{2} = (3510.7\pm0.7)\times 10^{6}$ and $J_{4} = (-33.61\pm 1.0)\times 10^{6}$ for Uranus, and $J_{2} = (3529.4\pm4.5)\times 10^{6}$ and $J_{4} = (-35.8\pm 2.9)\times 10^{6}$ for Neptune (\citet{helled2010} and \citet{nett2013} as based on \citet{jacobson2009} and \citet{lindal1992}). These observed values are shown in dark pink in Figure \ref{fig:Js}. However, these values are influenced by the zonal winds in the dynamical atmospheres of Uranus and Neptune, whilst here we only compute the static contributions. \citet{kaspi2013} constrained the depth of winds by decomposing the gravity harmonics into a static and a dynamical contribution. The latter arises from perturbations to the density by the winds, which influence the gravity field. They estimated the dynamical perturbation by comparing their $J_4$ static values from wind-free interior structure models with the observed $J_4$. Therefore, in Figure \ref{fig:Js} we also show wind-corrected static values in light pink by subtracting the dynamic contributions obtained by \cite{kaspi2013} from the observed values. Table \ref{tab:j2n} summarises the observed and wind-corrected $J_2$ and $J_4$ values we used.
    While for Jupiter the wind-correction is tightly constrained thanks to the measurement of odd harmonics \citep{kaspi2018} and leads to a lower static $|J_4|$ value, for Uranus and Neptune the wind-correction is solely inferred from the possible range obtained from interior models, and at present, it acts to enlarge the uncertainty in the static values.

\begin{table}[ht]
    \begin{threeparttable}
        \centering
        \small
        \caption{Observed, wind-corrected gravitational harmonics $J_2$ and $J_4$ and dynamical correction.}
        
            \begin{tabular}{c|cc|c}
            Description& $J_2/10^{-6}$& $J_4/10^{-6}$&Reference\\\hline\hline
            observed-U&3510.7$\pm$0.7&-33.61$\pm$ 1&\citet{helled2010}\\
            observed-N&3529.4$\pm$4.5 &-35.80$\pm$ 2.9&\citet{helled2010}\\
            dyn. corr.-U&-&-1$\leq\Delta J_4\leq$+3&\citet{kaspi2013}\\
            dyn. corr.-N&-&-5$\leq\Delta J_4\leq$+4&\citet{kaspi2013}\\
            static-U&3510.7$\pm$0.7&-32.61 to -36.61&this work \\
            static-N&3529.4$\pm$4.5&-30.80 to -39.80&this work \\\hline
            \end{tabular}
        
        \begin{tablenotes}
            \item \textbf{Notes.} Values are for $R_{\rm eq}$ at 1 bar, as used by \citet{helled2010} and \citet{nett2013} based on \citet{lindal1992} and \citet{jacobson2009}. Wind corrections from \citet{kaspi2013} are stated and the range of corrected static $J_4$ shown in bright pink in Figure \ref{fig:Js} are given. 
        \end{tablenotes}
    \end{threeparttable}
 
\label{tab:j2n}
\end{table}

\subsection{Results}\label{sec:structure_results}

\subsubsection{Gravitational harmonics}

\begin{figure*}[!tbp]
  \centering
  \includegraphics[width=1.0\textwidth]{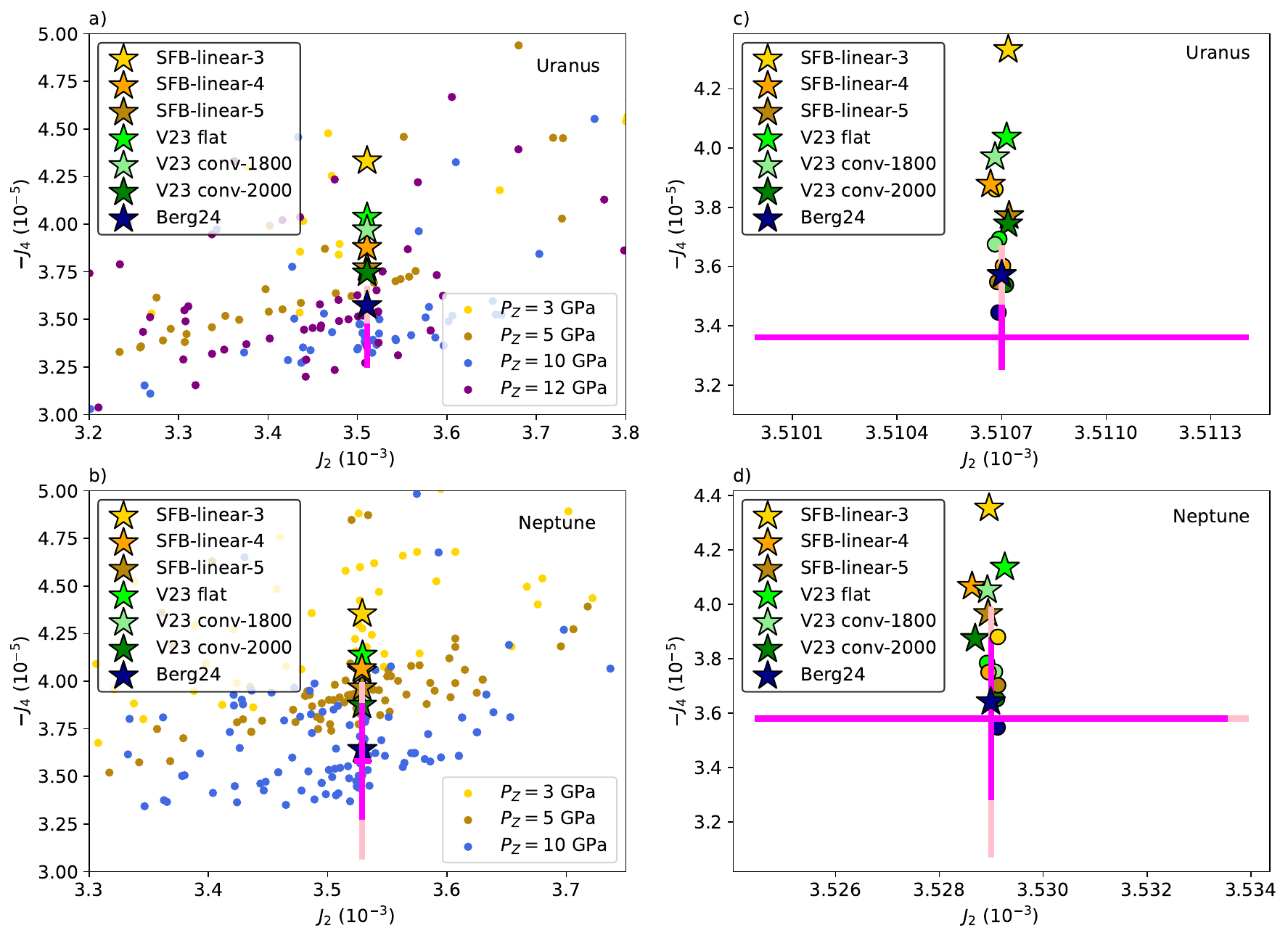}
  \caption{Gravitational harmonics $J_{2}$, $J_{4}$ for Uranus (a) and Neptune (b) in case (ii) models (stars) and case (iv) models (dots). The colour of the dots indicates the parameter $P_Z$.  The colour of the stars indicates the H$_2$-H$_2$O phase diagram used. 
  Dark-pink crosses show observed values. Light-pink crosses are wind-corrected static $J_{4}$ values (see text for details). In the case of Neptune, the V23 conv-1800 K 
  and SFB-linear-4 GPa stars overlap. Note the change in units for  $J_{2}$ and $J_{4}$ with respect to Table \ref{tab:j2n}. Panels (c) and (d) show stars: same models as in (a) and (b), and circles: case (iii) models. This figure shows that a mixture of rocks (circles) in the deep interior below the 2000 K level, and water reduces the resulting $|J_4|$ values. }
  \label{fig:Js}
\end{figure*}

    In Figures \ref{fig:Js}a and \ref{fig:Js}b we plot the gravitational harmonics $J_2$, $J_4$ of our seven case (ii) models (star symbols) and of several hundred case (iv) models for Uranus and Neptune with randomly selected parameters (dots). Figures \ref{fig:Js}c and \ref{fig:Js}d show a zoom-in to compare case (ii) models (stars) with case (iii) models (circles). 

    The choice of $P_Z$ (colour-coded) in the case (iv) models imposes a diagonal lower limit in the $|J_4|$-$J_2$ plane. This lower limit decreases with increasing $P_Z$. This means that deeper water-poor to water-rich transitions reduce the $|J_4|$ value at a given $J_2$, a behaviour that is well-known for Jupiter models \citep{nettelmann2011}. Above that lower limit, the solutions for different $P_Z$ overlap. An upper limit is expected to occur the more homogeneous the planet is assumed to be (i.e. for higher values of $Z_1$). However, we have not explored the full parameter space far away from the observed $J_4$ values. 

    For Uranus, case (iv) models suggest that if a sharp compositional boundary exists, it should be deeper than 3 GPa, as the models for 3 GPa lie above the $J_4$ value, wind-corrected or not. Models with $P_Z$ at 5 GPa (brown) cross the upper limit of the wind-corrected $J_4$ values. However, to reach the ultimate lower limit of the wind-corrected $|J_4|$, water-poor to water-rich transitions deeper than 10 GPa would be needed. For Neptune, the much larger observational uncertainty in $J_4$ permits models with $P_Z=5$ GPa. 

    This behaviour with $P_Z$ is well reflected in the three case (ii) models with cut-off pressures at 3, 4, and 5 GPa: while the SFB-linear-3 GPa Uranus model (yellow star) is outside the uncertainty range in $J_4$, the SFB-linear-5 GPa model (brown star) is close to the upper limit of $|J_4|$. 
    
    The deepest $P_Z$ is obtained for the Berg24-constrained model. Its $J_4$ values for Uranus and Neptune are well within the uncertainty range. Thus the Berg24 model (blue star) permits a classical ice giant interior, where the water-poor to water-rich transition is caused by H$_2$-H$_2$O phase separation. 

    In the case (ii) series, and in agreement with previous three-layer structure models \citep{nett2013, Bailey_2021}, we find a water-rich deep interior with $Z_\text{2} = Z_{\rm H_{2}O}$ ranging from [0.68--0.87] for Uranus and [0.73--0.90] for Neptune. 
    These results show that models where the atmospheric water abundances and the transition pressures are constrained by H$_2$-H$_2$O phase separation match the known gravity field of Uranus if $P_Z \gtrsim 10$ GPa, and for Neptune if $P_Z \gtrsim 5$ GPa. The constraints and resulting $J_{4}$ values of the case (ii) star models are listed in Table \ref{tab:ZatmPZJ4}.

\begin{table*}[ht]
    \begin{threeparttable}
        \centering
        \caption{Case (ii) and case (iii) constrained models with resulting $J_{4}$ values and required water and rock abundances and core mass.}
            \begin{tabular}{cc|ccccc|cccc}
                Planet & Phase & $\Zatm$ & $P_Z$ &  $Z_{2,\rm H_{2}O}$ & $J_4 / 10^{-6}$ & $M_{\mathrm{core}}$ & $Z_{2,\rm H_{2}O}$ & $Z_{2,\rm rocks}$ &  $J_4/10^{-6}$ & $M_{\mathrm{core}}$  \\
                 & diagram & & (GPa) & case (ii) & case (ii) & $(M_{\mathrm{E}})$&case (iii) & case (iii) & case (iii) & $(M_{\mathrm{E}})$ \\\hline\hline
                U & SFB-linear-3 GPa & 0.209 & 3 & 0.683 & -43.32 &4.134& 0.437&0.324 & -38.61 & 2.252 \\ 
                U & SFB-linear-4 GPa & 0.058 & 4 & 0.775 & -38.78 &2.953&0.449& 0.332 & -36.01 &1.777\\ 
                U & SFB-linear-5 GPa & 0.057 & 5 & 0.799 & -37.71 &2.582 &0.455& 0.337 & -35.48 &1.556\\ 
                U & V23 conv-1800 K & 0.133 & 4.8 & 0.759 & -39.70 &3.165 &0.447 &0.331 & -36.75 &1.870\\ 
                U & V23 conv-2000 K & 0.085 & 6 & 0.808 & -37.43 &2.431&0.458& 0.339 & -35.38 &1.428\\ 
                U & V23 flat & 0.114 & 3.7 & 0.743 & -40.36 &3.397& 0.443&0.328 & -36.93 &1.989\\ 
                U & Berg24  & 0.141 & 11 & 0.864 & -35.72 &1.467&0.479& 0.355 & -34.45 &0.654 \\\hline
                N & SFB-linear-3 GPa & 0.159 & 3 & 0.734 & -43.54 &5.389&0.454& 0.336 & -38.80 &3.332\\ 
                N & SFB-linear-4 GPa & 0.058 & 4 & 0.793 & -40.66 &4.517& 0.460&0.341 & -37.50 &3.057\\ 
                N & SFB-linear-5 GPa & 0.057 & 5 & 0.814 & -39.66&4.142 &0.465& 0.344 & -37.02&2.838\\ 
                N & V23 conv-1800 K & 0.094 & 4.8 & 0.796 & -40.53&4.439 &0.461& 0.341 & -37.52&2.996\\ 
                N & V23 conv-2000 K & 0.057 & 6 & 0.834 & -38.74 &3.771& 0.471&0.349 & -36.51&2.591\\ 
                N & V23 flat & 0.078 & 3.7 & 0.778 & -41.36 &4.756& 0.457&0.339 & -37.84 &3.162\\ 
                N & Berg24 & 0.101 & 11 & 0.892 & -36.41 &2.524&0.635*& 0.235* & -35.47 &1.880\\\hline
            \end{tabular}
        \begin{tablenotes}
            \item \textbf{Notes.} In case (ii) the heavy element is only water whereas in case (iii) rocks are added. $Z_{2,\rm H_{2}O}$ is obtained as $Z_{2,\rm rocks}$ $\times$ I/R. *I/R=1x solar.
        \end{tablenotes}
    \end{threeparttable}
 
\label{tab:ZatmPZJ4}
\end{table*}

    The classic models of the ice giants with interiors highly enriched in water and with a high I:R ratio may not be realistic. In the case (iii) series of models we investigate the effect on $J_4$ of adding rocks to the deep envelope. We set the I:R factor to 0.5 times the solar value (or 1x) and vary $Z_{\rm 2,H_2O}$, and consequently $Z_{2,\rm rocks}$. Exploring the full range of I:R factors is out of the scope of this work.
    Figure \ref{fig:Js}c and Figure \ref{fig:Js}d display such models (circles) together with the seven stars of case (ii). The stronger central condensation of mass in case (iii) models tends to reduce $|J_4|$ and to provide a better fit. Moreover, due to the presence of rocks, a lower water abundance in the deep interior is needed. In these case (iii) constrained models, we find solutions with $Z_{\rm 2,H_2O}$ in the range [0.44--0.48] and $Z_{\rm 2,rocks}$ in the range  [0.32--0.36] for Uranus, while $Z_{\rm 2,H_2O}$ = [0.45--0.64] and $Z_{2,\rm rocks}$ = [0.34--0.24] for Neptune.

\subsubsection{Adiabatic gradient}

\begin{figure}[!tbp]
  \centering
  \includegraphics[width=0.5\textwidth]{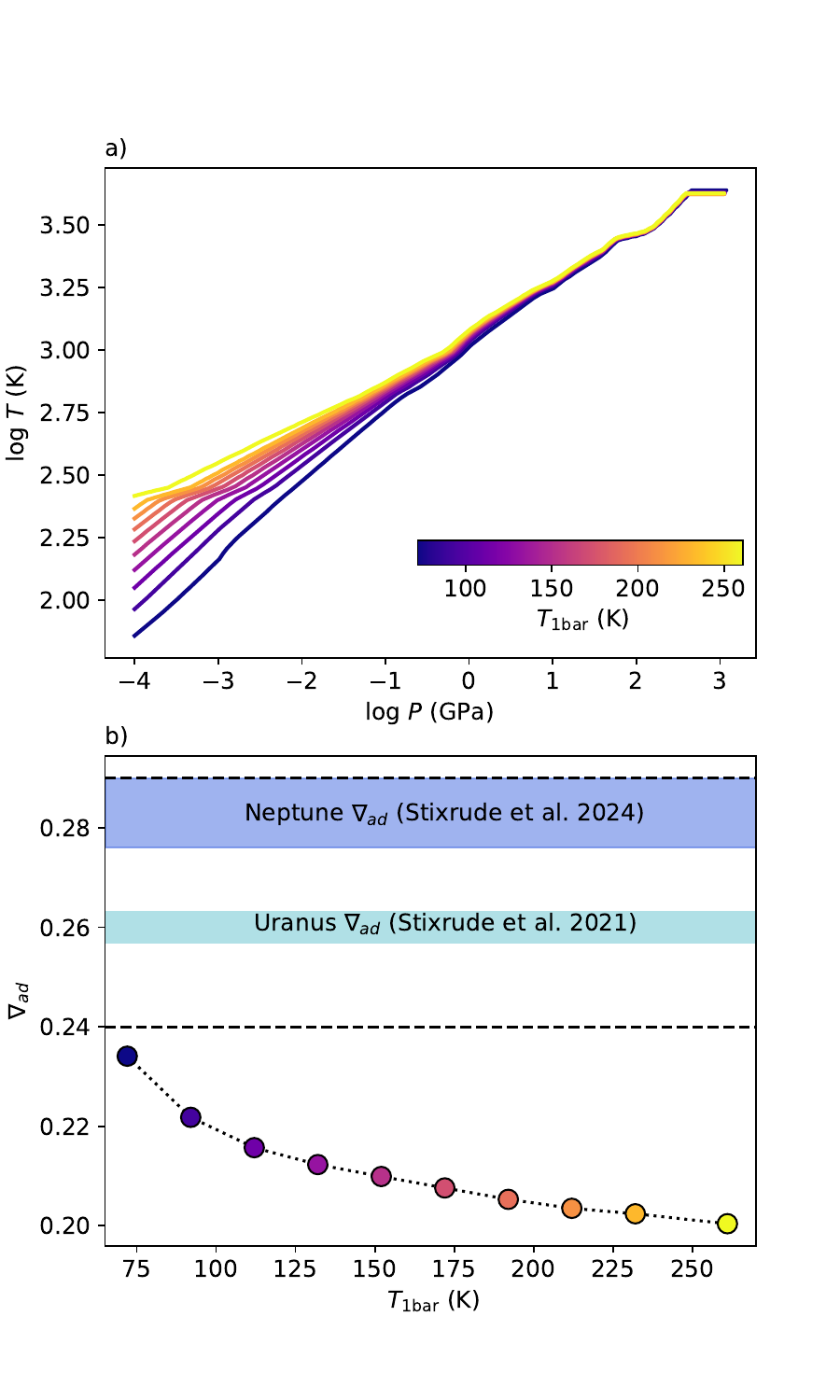}
  \caption{Comparison of adiabatic gradient obtained from models with phase separation with values used by \cite{stix21} and \cite{stix24}.(a): Interior $P-T$ profiles for Neptune for different $\Tone$ temperatures. The metallicity $Z_1 = \Zatm$ is varied with $\Tone$ according to the Berg24 phase diagram while $Z_2$ essentially stays at $\sim 0.7$. (b): Bulk volume-weighted-adiabatic gradient over temperature since the start of demixing (right side of the x-axis) to the present temperature. Blue bands show the range of values for $\nabla_{\rm ad}$ found by \citet{stix21} for Uranus and \citet{stix24} for Neptune to match the cooling times, amongst their range of tested values between 0.24 and 0.29 (dashed lines).}
  \label{fig:grad_ad}
\end{figure}

    Phase separation and the associated change in the abundances will affect the planetary $P$--$T$ profile and thus the adiabatic temperature gradient, $\nabla_{\rm ad}$, even if the planet remains adiabatic. Recently, \citet{stix21} and \citet{stix24} considered the possibility of Uranus and Neptune having a growing frozen core, whose size is determined by the transition between the fluid and superionic phase of water. The authors study the influence of this effect on the thermal evolution and tidal dissipation of the two ice giants. For their interior models, they test a range of adiabatic gradients and specific heat but keep their values constant throughout the interior and the evolution. They show that a frozen core growing over time can explain the observed luminosity of Uranus and Neptune, as well as a time-varying tidal dissipation that can explain the evolution/migration of their satellites. They also find that a slightly different range of adiabatic gradient values (and heat capacities) between both planets (see Figure \ref{fig:grad_ad}) can match the luminosities of each planet and associate this difference to a difference in composition between the planets \citep{stix24}. 

    In Figure \ref{fig:grad_ad} we show adiabatic $P$--$T$ profiles for various 1 bar temperatures that occur during the evolution of the ice giants, and the associated bulk-volume-weighted adiabatic gradient ($\nabla_\text{ad}$) of Neptune. In the underlying structure models, the water abundances in the outer envelopes change due to phase separation. 
    We find that our bulk volumetric $\nabla_\text{ad}$ value for present Neptune is lower than the range considered by \citet{stix24}. Our lower values result partially from the AQUA-EOS, which yields colder adiabats than H2O-REOS. Moreover, we find that $\nabla_\text{ad}$ changes by about $0.03$ over the course of the evolution. This change with time is much larger than the uncertainty range that \citet{stix24} predict in order to fit the luminosities.
    We also find that $\nabla_\text{ad}$ changes within the interior, with larger values further out and smaller values deeper inside. Our results suggest that if $\nabla_\text{ad}$ is used to fit luminosity to infer internal composition, the full composition and time-dependent profile should be used because the change over time will influence the cooling behaviour.

\section{Discussion}\label{Discussion}

\subsection{$\Zatm$ from H$_2$-H$_2$O phase separation versus from models constrained by gravitational harmonics}

    In our models constrained by H$_2$-H$_2$O phase separation, $\Zatm$ is the equilibrium water abundance, which is adopted by the entire outer envelope, i.e. $\Zatm = Z_{A} = Z_{1}$. In contrast, in classical adiabatic structure models, $\Zatm=Z_{1}$ is determined by the fit to the gravitational harmonics. For our models constrained by phase diagrams only, we find that $\Zatm$ is necessarily smaller for Neptune than for Uranus simply because Neptune's observed $\Tone$ is $\sim 4\:$ K lower than Uranus' $\Tone$, and we extend the temperature profile adiabatically from the 1 bar level down to the phase separation region. 
    In classical three-layer structure models \citep{nett2013}, $Z$ in the outer envelope ($Z_{1}$) spans a rather narrow range of $0-0.2$ for Uranus and a wider range of  $0-0.6$ for Neptune. The higher uncertainty in Neptune's $Z_{1}$ is mainly due to the higher uncertainty in $J_2$ and $J_4$ than for Uranus \citep{nett2013,HelledFortney20}. Thus Neptune could have a lower $Z_{1}$ than Uranus in agreement with structure models.  However, for the majority of structure models, Neptune's $Z_1$ is larger than that of Uranus.  \citet{Bailey_2021} even find an envelope water mole fraction of $\sim 0.01$ for all of their Uranus models and of $0.15-0.20$ for all of their Neptune models, thus also systematically higher $\Zatm$ values for Neptune than for Uranus. 

    It is unknown whether or not $Z_{1}$ in Neptune is higher than in Uranus. If it was indeed higher, and if this $Z_{1}$ was due to water, our models would imply that the $P-T$ profiles do not extend adiabatically downward. Deviation from adiabaticity can arise from inhibition of convection across the methane or the water cloud layer if their abundances are high enough \citep{Guillo95,Leconte17,MarkhamStevenson21}, in which case a superadiabatic temperature gradient develops. If that gradient was larger in Neptune than in Uranus, perhaps because of the higher heat flow, Neptune's deep adiabat could be warmer than that of Uranus. As a consequence, phase-diagram-constrained models could then predict a higher $Z_{1}$ for Neptune than for Uranus, and in both cases, this $Z_{1}$ would differ from the measurable values in their atmospheres due to inhibited convection.

    It is also possible that latent heat release from methane or water condensation leads to a colder deep adiabat than seen at the 1-bar level \citep{Kurosaki17,MarkhamStevenson21}. This effect would have to be stronger in Uranus to reduce the temperatures along the deep adiabat more than in Neptune.

    Furthermore, with our phase separation model we only address the water abundance, while in the actual planets, $\Zatm$ is likely composed of methane and other condensable species in addition to water. Our models could then suggest that a higher $Z_1$ in Neptune's structure models is due to methane rather than water, with Neptune's methane abundance being larger than Uranus'. However, such a scenario is not supported by current observational data, which find $80\pm 20$ solar C/H enrichment in both planets' atmospheres \citep{Atreya20}.

\subsection{Mixing with other elements}

    Our $\Zatm$ values are computed for H-O mixtures according to Equation \ref{eq:xZconversion}. In the presence of helium or carbon, potentially the most abundant other elements in the atmosphere, the water mass fraction would decrease, for a given $\xWater$/H ratio. Thus our models predict upper limits on the water mass fraction. These are $Z_1=0.21$ for Uranus and $Z_1=0.16$ for Neptune. These upper limits are relevant for the determination of the pressure at the bottom of the water clouds, and the vertical water abundance profile across the cloud up to the top, where the water abundance may be measurable by an entry probe or remote sensing from an orbiter, like Juno at Jupiter.

\subsection{Demixing of other elements}

    Demixing may also occur among other elements in Uranus and Neptune and shape their structures.
     H-He demixing occurs at Mbar pressures \citep{SR18}. In Jupiter, this is the only accepted explanation for its atmospheric He-depletion and even stronger Ne-depletion \citep{zahn1998, stevenson1998, wilson&mil2010}. In Saturn, H-He demixing has also been proposed to explain its high luminosity \citep{fortney2003, PUSTOW2016} and co-axial magnetic field, although there are alternative explanations \citep{LC13}. H-He demixing in Saturn is predicted by all recent theoretical H-He phase diagrams. As the interior of the ice giants is likely even colder than that of the much more massive gas giants, H-He phase separation may also occur in Uranus and Neptune unless their deep interiors do not contain a H-He gas or the deep interior is shielded by a strong thermal boundary layer (TBL) \citep{scheibe21,nett24}, which may result from the water-poor to water-rich transition \citep{nettelmann16,scheibe21} or from the reduced convection in a frozen core \citep{stix21,stix24}.

    A strong TBL would also influence possible demixing between H and C, which diamond-anvil-cell experiments find to occur at pressures of $\sim 30$ GPa \citep{watkins22}, while in laser-shock-compression experiments it is seen not before Mbar pressures are reached \citep{Kraus17} although the presence of water appears to support diamond formation at intermediate pressures \citep{zhe22}. C-H demixing will act to enrich the atmosphere in H, while C would form diamonds that sink downward. The process of C-H demixing could leave behind a water-rich envelope as assumed in many three-layer models. At present it is still unclear how mixtures of H-He-C-O behave at high pressures.

    Further elements which may shape the structure of the ice giants and perhaps even more so of warm Neptune-like planets, are Mg and Si from the rock-forming refractory elements. Mixing is experimentally seen in the Mg-O-H system under ice giant conditions around 20--30 GPa and 2000 K \citep{kim2021}, while in numerical simulations mixing is seen around 174 GPa and 6000 K \citep{kova2023}. It is clear that further studies on the (de)mixing behaviour in H-He-CNO-Mg-Si systems are needed to advance our understanding of Neptune-like planets. 

\subsection{A gradual transition?}

    We find that the water-poor/water-rich transition is sharp because all our constructed phase diagrams assume that the shape of the demixing curve $\Tdmx(\xWater)$ at 0.2 GPa also holds at higher pressures. In contrast, isobaric H-He demixing curves have a more complex shape (see the phase diagram by \citep{SR18} so that phases of moderate compositional differences (He-enriched, He-depleted) are possible whose composition is highly sensitive to temperature \citep{chang2024}. Should such behaviour also apply to the O-H system at pressures of a few GPa rather than the flat shape we adopted (see Figure \ref{fig:phase} and Section 2.3) one would also expect a gradual change in the water abundance in the ice giants. Nevertheless, the region of pressures between 4--10 GPa is narrow, and therefore a potential gradual transition would be much narrower than the He-rain region in gas planets, where it may extend over several hundred GPa \citep{nett15,chang2024,HowardHelledAA24}. Whether or not a gradual transition is possible that suppresses convection and leads to a warmer interior is at present unknown.

\subsection{What if no phase separation occurs?}

    Let us suppose H$_2$-H$_2$O phase separation does not occur in the ice giants. The low-density outer envelope in Uranus found in structure models would then require a different explanation. Formation models predict a gradual decrease of the heavy element abundance in the accreted material with time, leading to an outward decreasing $Z$ \citep{Helled17}. However, formation models do not predict a sharp transition from water-poor to water-rich at about 0.85-0.9 of the ice giants' radius unless a very specific formation path is assumed \citep{VenturiniHelled17}. We may thus assume that the water abundance extends homogeneously down through the atmosphere before it gradually increases, with water being mixed with rocks deeper in the planet. 
    Furthermore, an extended homogeneous, convective and electrically conductive region such as the ionic water region is required for a dynamo to operate \citep{stanley2004, stanley2006}. Therefore, in a fully inhomogeneous planet where a $Z$-gradient inhibits convection, it is unclear how the observed magnetic field could be generated. 
    If the low water abundance found for the outer envelope extended deeper down, then another candidate for a sharp increase in $Z$ would be needed. This could be due to rock clouds at about 2000 K and 20 GPa. Alternatively, rocky material may be confined to the deeper layers because accreted rocky planetesimals ablate less efficiently than icy ones. This could lead to a compositional rock-gradient.

\subsection{What do formation models predict?}

    Assessing our atmospheric water abundance estimates in the frame of formation models is somewhat entangled given a) the constraints to these models are based on interior structure model estimates, and b) the various hypotheses of formation currently being investigated. The standard scenario of core accretion \citep{pollack1996} seems to support the heavy element amount thought to be in the ice giants, but the in-situ formation timescales are much longer than planetary disk lifetimes. Although this timescale issue can be surpassed using higher accretion rates, matching the constraints from structure models, more specifically the H/He and heavy element estimates, becomes the issue as shown by \cite{helled2014}, as well as explaining why Uranus and Neptune did not undergo runaway gas accretion and became gas giants. This all leads to the need for very particular conditions to meet the available requirements, hence a fine-tuning issue. 
    
    In the frame of formation models based on pebble accretion, following the pebble accretion rate from \cite{lambrechts14}, \cite{vallettahelled2022} find that Uranus and Neptune may have been formed in situ within the lifetime of protostellar disks. To validate their models, they focus on Uranus and Neptune accreting H/He envelopes as estimated by structure models by \cite{helled11} and \cite{nett2013}. \cite{vallettahelled2022} then obtain certain formation scenarios that match the timescales and the H/He mass, and also have a low outer envelope heavy element metallicity (in their case this is pure water) of 0.03 in mass fraction (see  Figure \ref{fig:zatm}). Their estimate is not far from our water-depleted atmosphere estimates. \cite{vallettahelled2022} find however that apart from these successful cases, in many others, the planets must have accreted an extra amount of heavy elements after their formation (through giant impacts, for example) to account for the missing heavy elements inside.
    
    Moreover, \cite{Mousis24} use the high carbon enrichment observed in the atmospheres of Uranus and Neptune \citep{Atreya20} and the suggestion of \cite{ali-dib2014} on the possible formation of the ice giants at the CO line to estimate the evolution and abundances of certain species in this region. We converted their estimates of bulk envelope oxygen enrichments with respect to protosolar. These are also shown in Figure \ref{fig:zatm}. Given that these are bulk estimates, and therefore different to our $\Zatm$, we don't compare our results to these estimates directly.  

\section{Summary and conclusions} \label{conclusion}

    Based on the presented results, we draw the following conclusions:

    \begin{itemize}
      \item We have constructed seven phase diagrams based on experimental data up 4 GPa and theoretical data up to 12 GPa that span the current level of uncertainty. In all cases, we obtain an overlap with the adiabats of Uranus and Neptune. This leads to a strong rain-out of water in the planets.
      \item Assuming no barrier between the atmosphere and the region of phase separation, we predict a low atmospheric water mass fraction abundance $\Zatm$ of 0.05--0.21 for Uranus, and of 0.05--0.16 for Neptune.
      \item The resulting water-poor to water-rich transition is sharp, and occurs between $P_Z = 4$ and 11 GPa, depending on phase diagram uncertainties.
      \item Structure models constrained by the $\Zatm$ and $P_Z$ from H$_2$-H$_2$O phase separation fit $J_2$. Whether or not the models also fit $J_4$ depends sensitively on $P_Z$. For phase diagrams with $P_Z \lesssim 4$ GPa, the resulting $|J_4|$ values are too large for both Uranus and Neptune, while for phase diagrams with $P_Z \gtrsim 10$ GPa (Uranus) or 5 GPa (Neptune), the upper limit of the wind-corrected $|J_4|$ values is reached. The demixing curve of \citet{berg24} leads to the deepest $P_Z$ and works for both planets. 
      \item Structure models with rocks below an assumed rock cloud condensation level at 2000 K (or deeper) have lower $|J_4|$ values, so that the observed mean values can be reached, suggesting that true low $|J_4|$ values could indicate a deeper rock-poor/rock-rich transition. A reduced uncertainty in the observed $J_4$ and the wind contribution of both Uranus and Neptune are needed to rule out or support the presented models. 
      \item The adiabatic gradient changes by $\sim 10\%$ over time. This change is expected to influence the cooling of the planet.
      \item The H-O system may not be the only relevant one that shapes the interior structure of the ice giants. In the C-H system, demixing has also been found both in experiments and in simulations. Different systems may have different separation locations, and effectively lead to gradual $Z$-transitions, where convection can be suppressed. 
      \item The sinking of water releases gravitational energy, and leads to the expansion of the then less dense outer envelope. While the effect is likely small, it should be quantified and compared to the low luminosity of Uranus in future work. 
      \item Future work using evolution models will provide a more complete picture of the consequences of this process on the internal structure and thermal evolution of the planets. 
    \end{itemize}

   The ice giant community is making efforts to design a space mission to explore the ice giants \citep[e.g.][]{fletcher2020}. The Decadal Survey \cite{decadal} emphasizes the many open questions regarding our understanding of their formation, interior structure and evolution and proposes a Uranus Orbiter and Probe (UOP) as NASA's next flagship mission with the highest priority. We of course hope that this will indeed become a reality and want to emphasise the direct relation between the work presented here and the decadal study question 7 (specifically 7.1 and 7.2). Obtaining new gravity field data would be crucial, and extremely helpful to set a constraint on the possible deep water abundance. Furthermore, in-situ abundance measurements from an atmospheric probe would result in further constraints on the volatile abundances, bringing us a step closer to understanding what the interior composition of these planets looks like.

\begin{acknowledgements}
We acknowledge the support of the German Science Foundation (DFG) through the research unit FOR 2440 “Matter under planetary interior conditions” (projects PA 3689/1-1 and NE 1734/2-2). N.N.~acknowledges support through NASA's Juno Participating Scientist Program under grant 80NSSC19K1286. We also thank the reviewer for their thorough feedback.
\end{acknowledgements}

%
%

\bibliographystyle{aa} 
\bibliography{bibliography}

\begin{thebibliography}{79}
\expandafter\ifx\csname natexlab\endcsname\relax\def\natexlab#1{#1}\fi

\bibitem[{Ali-Dib {et~al.}(2014)Ali-Dib, Mousis, Petit, \&
  Lunine}]{ali-dib2014}
Ali-Dib, M., Mousis, O., Petit, J.-M., \& Lunine, J.~I. 2014, The Astrophysical
  Journal, 793, 9

\bibitem[{Atreya {et~al.}(2020)Atreya, Hofstadter, In, Mousis, Reh, \&
  Wong}]{Atreya20}
Atreya, S., Hofstadter, M., In, J., {et~al.} 2020, SSRv, 216, 18

\bibitem[{Bailey \& Stevenson(2021)}]{Bailey_2021}
Bailey, E. \& Stevenson, D.~J. 2021, The Planetary Science Journal, 2, 64

\bibitem[{{Bali} {et~al.}(2013){Bali}, {Audétat}, \& {Keppler}}]{bali13}
{Bali}, E., {Audétat}, A., \& {Keppler}, H. 2013, Nature, 495

\bibitem[{Baumeister {et~al.}(2020)Baumeister, Padovan, Tosi, Montavon,
  Nettelmann, MacKenzie, \& Godolt}]{Baumeister_2020}
Baumeister, P., Padovan, S., Tosi, N., {et~al.} 2020, The Astrophysical
  Journal, 889, 42

\bibitem[{{Baumeister, Philipp} \& {Tosi, Nicola}(2023)}]{Baumeister_2023}
{Baumeister, Philipp} \& {Tosi, Nicola}. 2023, A\&A, 676, A106

\bibitem[{Becker {et~al.}(2014)Becker, Lorenzen, Fortney, Nettelmann,
  Schöttler, \& Redmer}]{Becker14}
Becker, A., Lorenzen, W., Fortney, J.~J., {et~al.} 2014, The Astrophysical
  Journal Supplement Series, 215, 21

\bibitem[{{Bergermann} {et~al.}(2021){Bergermann}, {French}, \&
  {Redmer}}]{berg21}
{Bergermann}, A., {French}, M., \& {Redmer}, R. 2021, Physical Chemistry
  Chemical Physics (Incorporating Faraday Transactions), 23, 12637

\bibitem[{Bergermann {et~al.}(2024)Bergermann, French, \& Redmer}]{berg24}
Bergermann, A., French, M., \& Redmer, R. 2024, Phys. Rev. B, 109, 174107

\bibitem[{Bolton {et~al.}(2017)Bolton, Lunine, Stevenson, Connerney, Levin,
  Owen, Bagenal, Gautier, Ingersoll, Orton, Guillot, Hubbard, Bloxham,
  Coradini, Stevens, Mokashi, Thorne, \& Thorpe}]{Bolton17}
Bolton, S., Lunine, J., Stevenson, D., {et~al.} 2017, Space Sci.~Rev, 213, 5

\bibitem[{Chabrier \& Debras(2021)}]{cd21}
Chabrier, G. \& Debras, F. 2021, The Astrophysical Journal, 917, 4

\bibitem[{Chabrier {et~al.}(2019)Chabrier, Mazevet, \& Soubiran}]{chabrier19}
Chabrier, G., Mazevet, S., \& Soubiran, F. 2019, The Astrophysical Journal,
  872, 51

\bibitem[{Chang {et~al.}(2024)Chang, Chen, Zeng, Wang, Chen, Tong, Yu, Kang,
  Zhang, Guo, Hou, Zhao, Yao, ma, \& Dai}]{chang2024}
Chang, X., Chen, B., Zeng, Q., {et~al.} 2024, Nature Communications, 15

\bibitem[{Fletcher {et~al.}(2020)Fletcher, Helled, Roussos, Jones, Charnoz,
  André, Andrews, Bannister, Bunce, Cavalié, Ferri, Fortney, Grassi, Griton,
  Hartogh, Hueso, Kaspi, Lamy, Masters, Melin, Moses, Mousis, Nettleman,
  Plainaki, Schmidt, Simon, Tobie, Tortora, Tosi, \& Turrini}]{fletcher2020}
Fletcher, L.~N., Helled, R., Roussos, E., {et~al.} 2020, Planetary and Space
  Science, 191, 105030

\bibitem[{Fortney \& Hubbard(2003)}]{fortney2003}
Fortney, J.~J. \& Hubbard, W.~B. 2003, Icarus, 164, 228–243

\bibitem[{{Guillot}(1995)}]{Guillo95}
{Guillot}, T. 1995, Science, 269, 1697

\bibitem[{{Haldemann, Jonas} {et~al.}(2020){Haldemann, Jonas}, {Alibert, Yann},
  {Mordasini, Christoph}, \& {Benz, Willy}}]{aqua2020}
{Haldemann, Jonas}, {Alibert, Yann}, {Mordasini, Christoph}, \& {Benz, Willy}.
  2020, A\&A, 643, A105

\bibitem[{He {et~al.}(2022)He, Rödel, Lütgert, Bergermann, Bethkenhagen,
  Chekrygina, Cowan, Descamps, French, Galtier, Gleason, Glenn, Glenzer,
  Inubushi, Hartley, Hernandez, Heuser, Humphries, Kamimura, \& Kraus}]{zhe22}
He, Z., Rödel, M., Lütgert, J., {et~al.} 2022, Science Advances, 8

\bibitem[{Helled {et~al.}(2010)Helled, Anderson, Podolak, \&
  Schubert}]{helled11}
Helled, R., Anderson, J.~D., Podolak, M., \& Schubert, G. 2010, The
  Astrophysical Journal, 726, 15

\bibitem[{{Helled} {et~al.}(2010){Helled}, {Anderson}, \&
  {Schubert}}]{helled2010}
{Helled}, R., {Anderson}, J.~D., \& {Schubert}, G. 2010, Icarus, 210, 446

\bibitem[{Helled \& Bodenheimer(2014)}]{helled2014}
Helled, R. \& Bodenheimer, P. 2014, The Astrophysical Journal, 789, 69

\bibitem[{{Helled} \& {Fortney}(2020)}]{HelledFortney20}
{Helled}, R. \& {Fortney}, J.~J. 2020, Philosophical Transactions of the Royal
  Society of London Series A, 378, 20190474

\bibitem[{Helled {et~al.}(2020)Helled, Nettelmann, \& Guillot}]{Helled_2020}
Helled, R., Nettelmann, N., \& Guillot, T. 2020, Space Science Reviews, 216

\bibitem[{Helled \& Stevenson(2017)}]{Helled17}
Helled, R. \& Stevenson, D. 2017, The Astrophysical Journal Letters, 840, L4

\bibitem[{{Howard, S.} {et~al.}(2024){Howard, S.}, {Müller, S.}, \& {Helled,
  R.}}]{HowardHelledAA24}
{Howard, S.}, {Müller, S.}, \& {Helled, R.} 2024, A\&A, 689, A15

\bibitem[{Hubbard \& Marley(1989)}]{hubbard1989}
Hubbard, W. \& Marley, M.~S. 1989, Icarus, 78, 102

\bibitem[{Hubbard \& MacFarlane(1980)}]{HMacF80}
Hubbard, W.~B. \& MacFarlane, J.~J. 1980, J. Geophys. Res., 88, 225

\bibitem[{Iess {et~al.}(2019)Iess, Militzer, Nicholson, Durante, Racioppa,
  Anabtawi, Galanti, Hubbard, Mariani, Tortora, Wahl, \& Zannoni}]{Iess19}
Iess, L., Militzer, Kaspi, Y., Nicholson, P., {et~al.} 2019, Science, 364, 2965

\bibitem[{{Jacobson}(2009)}]{jacobson2009}
{Jacobson}, R.~A. 2009, AJ, 137, 4322

\bibitem[{James \& Stixrude(2024)}]{stix24}
James, D. \& Stixrude, L. 2024, Space Science Reviews, 220

\bibitem[{Kaspi {et~al.}(2020)Kaspi, Galanti, Showman, Stevenson, Guillot,
  Iess, \& Bolton}]{kaspi2018}
Kaspi, Y., Galanti, E., Showman, A.~P., {et~al.} 2020, Space Science Reviews,
  216

\bibitem[{{Kaspi} {et~al.}(2013){Kaspi}, {Showman}, {Hubbard}, {Aharonson}, \&
  {Helled}}]{kaspi2013}
{Kaspi}, Y., {Showman}, A.~P., {Hubbard}, W.~B., {Aharonson}, O., \& {Helled},
  R. 2013, Nature, 497, 344

\bibitem[{{Kim} {et~al.}(2021){Kim}, {Chariton}, {Prakapenka}, {Pakhomova},
  {Liermann}, {Liu}, {Speziale}, {Shim}, \& {Lee}}]{kim2021}
{Kim}, T., {Chariton}, S., {Prakapenka}, V., {et~al.} 2021, Nature Astronomy,
  5, 815

\bibitem[{{Kova{\v{c}}evi{\'c}} {et~al.}(2023){Kova{\v{c}}evi{\'c}},
  {Gonz{\'a}lez-Cataldo}, \& {Militzer}}]{kova2023}
{Kova{\v{c}}evi{\'c}}, T., {Gonz{\'a}lez-Cataldo}, F., \& {Militzer}, B. 2023,
  Contributions to Plasma Physics, 63

\bibitem[{{Kraus} {et~al.}(2017){Kraus}, {Vorberger}, {Pak}, {Hartley},
  {Fletcher}, {Frydrych}, {Galtier}, {Gamboa}, {Gericke}, {Glenzer},
  {Granados}, {MacDonald}, {MacKinnon}, {McBride}, {Nam}, {Neumayer}, {Roth},
  {Saunders}, {Schuster}, {Sun}, {van Driel}, {D{\"o}ppner}, \&
  {Falcone}}]{Kraus17}
{Kraus}, D., {Vorberger}, J., {Pak}, A., {et~al.} 2017, Nature Astronomy, 1,
  606

\bibitem[{Kurosaki \& Ikoma(2017)}]{Kurosaki17}
Kurosaki, K. \& Ikoma, M. 2017, The Astronomical Journal, 153, 260

\bibitem[{{Lambrechts, M.} {et~al.}(2014){Lambrechts, M.}, {Johansen, A.}, \&
  {Morbidelli, A.}}]{lambrechts14}
{Lambrechts, M.}, {Johansen, A.}, \& {Morbidelli, A.} 2014, A\&A, 572, A35

\bibitem[{Leconte \& Chabrier(2013)}]{LC13}
Leconte, J. \& Chabrier, G. 2013, Nature Geoscience, 6, 347

\bibitem[{{Leconte, Jérémy} {et~al.}(2017){Leconte, Jérémy}, {Selsis,
  Franck}, {Hersant, Franck}, \& {Guillot, Tristan}}]{Leconte17}
{Leconte, Jérémy}, {Selsis, Franck}, {Hersant, Franck}, \& {Guillot,
  Tristan}. 2017, A\&A, 598, A98

\bibitem[{{Lindal}(1992)}]{lindal1992}
{Lindal}, G.~F. 1992, The Astronomical Journal, 103, 967

\bibitem[{Lindal {et~al.}(1987)Lindal, Lyons, Sweetnam, Eshleman, Hinson, \&
  Tyler}]{lindal1987}
Lindal, G.~F., Lyons, J.~R., Sweetnam, D., {et~al.} 1987, Journal of
  Geophysical Research, 92, 14987

\bibitem[{{MacKenzie, Jasmine} {et~al.}(2023){MacKenzie, Jasmine}, {Grenfell,
  John Lee}, {Baumeister, Philipp}, {Tosi, Nicola}, {Cabrera, Juan}, \& {Rauer,
  Heike}}]{MacKenzie_2023}
{MacKenzie, Jasmine}, {Grenfell, John Lee}, {Baumeister, Philipp}, {et~al.}
  2023, A\&A, 671, A65

\bibitem[{Malamud {et~al.}(2024)Malamud, Podolak, Podolak, \&
  Bodenheimer}]{malamud24}
Malamud, U., Podolak, M., Podolak, J.~I., \& Bodenheimer, P.~H. 2024, Icarus,
  421, 116217

\bibitem[{Mankovich {et~al.}(2016)Mankovich, Fortney, \& Moore}]{Mankovich16}
Mankovich, C., Fortney, J.~J., \& Moore, K.~L. 2016, The Astrophysical Journal,
  832, 113

\bibitem[{Markham \& Stevenson(2021)}]{MarkhamStevenson21}
Markham, S. \& Stevenson, D. 2021, The Planetary Science Journal, 2, 146

\bibitem[{Militzer {et~al.}(2016)Militzer, Soubiran, Wahl, \&
  Hubbard}]{militzer2016}
Militzer, B., Soubiran, F., Wahl, S.~M., \& Hubbard, W. 2016, Journal of
  Geophysical Research: Planets, 121, 1552

\bibitem[{Mousis {et~al.}(2024)Mousis, Schneeberger, Cavalié, Mandt,
  Aguichine, Lunine, Couzinou, Hue, \& Moreno}]{Mousis24}
Mousis, O., Schneeberger, A., Cavalié, T., {et~al.} 2024, The Planetary
  Science Journal, 5, 173

\bibitem[{Movshovitz {et~al.}(2020)Movshovitz, Fortney, Mankovich, Thorngren,
  \& Helled}]{mov2020}
Movshovitz, N., Fortney, J.~J., Mankovich, C., Thorngren, D., \& Helled, R.
  2020, The Astrophysical Journal, 891, 109

\bibitem[{{National Academies of Sciences Engineering and
  Medicine}(2023)}]{decadal}
{National Academies of Sciences Engineering and Medicine}. 2023, Origins,
  Worlds, and Life: A Decadal Strategy for Planetary Science and Astrobiology
  2023-2032 (Washington, DC: The National Academies Press)

\bibitem[{{Nettelmann}(2011)}]{nettelmann2011}
{Nettelmann}, N. 2011, \apss, 336, 47

\bibitem[{{Nettelmann} {et~al.}(2012){Nettelmann}, {Becker}, {Holst}, \&
  {Redmer}}]{nett2012}
{Nettelmann}, N., {Becker}, A., {Holst}, B., \& {Redmer}, R. 2012, ApJ, 750, 52

\bibitem[{{Nettelmann} {et~al.}(2024){Nettelmann}, {Cano Amoros}, {Tosi},
  {Fortney}, \& {Helled}}]{nett24}
{Nettelmann}, N., {Cano Amoros}, M., {Tosi}, N., {Fortney}, J.~J., \& {Helled},
  R. 2024, Space Science Reviews, 220, 56

\bibitem[{Nettelmann {et~al.}(2015)Nettelmann, Fortney, Moore, \&
  Mankovich}]{nett15}
Nettelmann, N., Fortney, J.~J., Moore, K., \& Mankovich, C. 2015, Monthly
  Notices of the Royal Astronomical Society, 447, 3422–3441

\bibitem[{Nettelmann {et~al.}(2013)Nettelmann, Helled, Fortney, \&
  Redmer}]{nett2013}
Nettelmann, N., Helled, R., Fortney, J., \& Redmer, R. 2013, Planetary and
  Space Science, 77, 143

\bibitem[{Nettelmann {et~al.}(2016)Nettelmann, Wang, Fortney, Hamel,
  Yellamilli, Bethkenhagen, \& Redmer}]{nettelmann16}
Nettelmann, N., Wang, K., Fortney, J., {et~al.} 2016, Icarus, 275, 107–116

\bibitem[{{Neuenschwander} \& {Helled}(2022)}]{neu22}
{Neuenschwander}, B.~A. \& {Helled}, R. 2022, \mnras, 512, 3124

\bibitem[{{Neuenschwander} {et~al.}(2024){Neuenschwander}, {M{\"u}ller}, \&
  {Helled}}]{neu24}
{Neuenschwander}, B.~A., {M{\"u}ller}, S., \& {Helled}, R. 2024, \aap, 684,
  A191

\bibitem[{Pearl {et~al.}(1990)Pearl, Conrath, Hanel, Pirraglia, \&
  Coustenis}]{PEARL1990}
Pearl, J., Conrath, B., Hanel, R., Pirraglia, J., \& Coustenis, A. 1990,
  Icarus, 84, 12

\bibitem[{Pearl \& Conrath(1991)}]{pearl&conrath}
Pearl, J.~C. \& Conrath, B.~J. 1991, Journal of Geophysical Research: Space
  Physics, 96, 18921

\bibitem[{Podolak {et~al.}(2022)Podolak, Malamud, \& Podolak}]{podolak22}
Podolak, J., Malamud, U., \& Podolak, M. 2022, Icarus, 382, 115017

\bibitem[{{Pollack} {et~al.}(1996){Pollack}, {Hubickyj}, {Bodenheimer},
  {Lissauer}, {Podolak}, \& {Greenzweig}}]{pollack1996}
{Pollack}, J.~B., {Hubickyj}, O., {Bodenheimer}, P., {et~al.} 1996, Icarus,
  124, 62

\bibitem[{Püstow {et~al.}(2016)Püstow, Nettelmann, Lorenzen, \&
  Redmer}]{PUSTOW2016}
Püstow, R., Nettelmann, N., Lorenzen, W., \& Redmer, R. 2016, Icarus, 267, 323

\bibitem[{Scheibe {et~al.}(2021)Scheibe, Nettelmann, \& Redmer}]{scheibe21}
Scheibe, L., Nettelmann, N., \& Redmer, R. 2021, A\&A, 650, A200

\bibitem[{{Scheibe, Ludwig} {et~al.}(2019){Scheibe, Ludwig}, {Nettelmann,
  Nadine}, \& {Redmer, Ronald}}]{scheibe19}
{Scheibe, Ludwig}, {Nettelmann, Nadine}, \& {Redmer, Ronald}. 2019, A\&A, 632,
  A70

\bibitem[{Sch\"ottler \& Redmer(2018)}]{SR18}
Sch\"ottler, M. \& Redmer, R. 2018, Phys. Rev. Lett., 120, 115703

\bibitem[{Seward \& Franck(1981)}]{Seward1981}
Seward, T. \& Franck, E. 1981, Berichte der Bunsengesellschaft f{\"u}r
  physikalische Chemie, 85, 2

\bibitem[{{Stanley} \& {Bloxham}(2004)}]{stanley2004}
{Stanley}, S. \& {Bloxham}, J. 2004, Nature, 428, 151

\bibitem[{{Stanley} \& {Bloxham}(2006)}]{stanley2006}
{Stanley}, S. \& {Bloxham}, J. 2006, Icarus, 184, 556

\bibitem[{{Stevenson}(1998)}]{stevenson1998}
{Stevenson}, D.~J. 1998, Journal of Physics Condensed Matter, 10, 11227

\bibitem[{Stixrude {et~al.}(2021)Stixrude, Baroni, \& Grasselli}]{stix21}
Stixrude, L., Baroni, S., \& Grasselli, F. 2021, The Planetary Science Journal,
  2, 222

\bibitem[{Valletta \& Helled(2022)}]{vallettahelled2022}
Valletta, C. \& Helled, R. 2022, The Astrophysical Journal, 931, 21

\bibitem[{Vazan {et~al.}(2022)Vazan, Sari, \& Kessel}]{Vazan22}
Vazan, A., Sari, R., \& Kessel, R. 2022, The Astrophysical Journal, 926, 150

\bibitem[{Venturini \& Helled(2017)}]{VenturiniHelled17}
Venturini, J. \& Helled, R. 2017, The Astrophysical Journal, 848, 95

\bibitem[{Visscher \& Moses(2011)}]{VissherMoses11}
Visscher, C. \& Moses, J.~I. 2011, The Astrophysical Journal, 738, 72

\bibitem[{{Vlasov} {et~al.}(2023){Vlasov}, {Aud{\'e}tat}, \&
  {Keppler}}]{Vlasov23}
{Vlasov}, K., {Aud{\'e}tat}, A., \& {Keppler}, H. 2023, Contributions to
  Mineralogy and Petrology, 178, 36

\bibitem[{von Zahn {et~al.}(1998)von Zahn, Hunten, \& Lehmacher}]{zahn1998}
von Zahn, U., Hunten, D.~M., \& Lehmacher, G. 1998, Journal of Geophysical
  Research: Planets, 103, 22815

\bibitem[{{Watkins} {et~al.}(2022){Watkins}, {Huber}, {Childs}, {Salamat},
  {Pigott}, {Chow}, {Xiao}, \& {Coe}}]{watkins22}
{Watkins}, E.~B., {Huber}, R.~C., {Childs}, C.~M., {et~al.} 2022, Scientific
  Reports, 12, 631

\bibitem[{Wilson \& Militzer(2010)}]{wilson&mil2010}
Wilson, H.~F. \& Militzer, B. 2010, Phys. Rev. Lett., 104, 121101

\bibitem[{{Zharkov} \& {Trubitsyn}(1978)}]{zharkov78}
{Zharkov}, V.~N. \& {Trubitsyn}, V.~P. 1978, {Physics of planetary interiors}
  (Pachart Pub House)

\end{thebibliography}

\begin{appendix}

\section{Response of $J_2$, $J_4$ to parameter variation}

\begin{figure}[!tbp]
  \centering
  \includegraphics[width=0.4\textwidth]{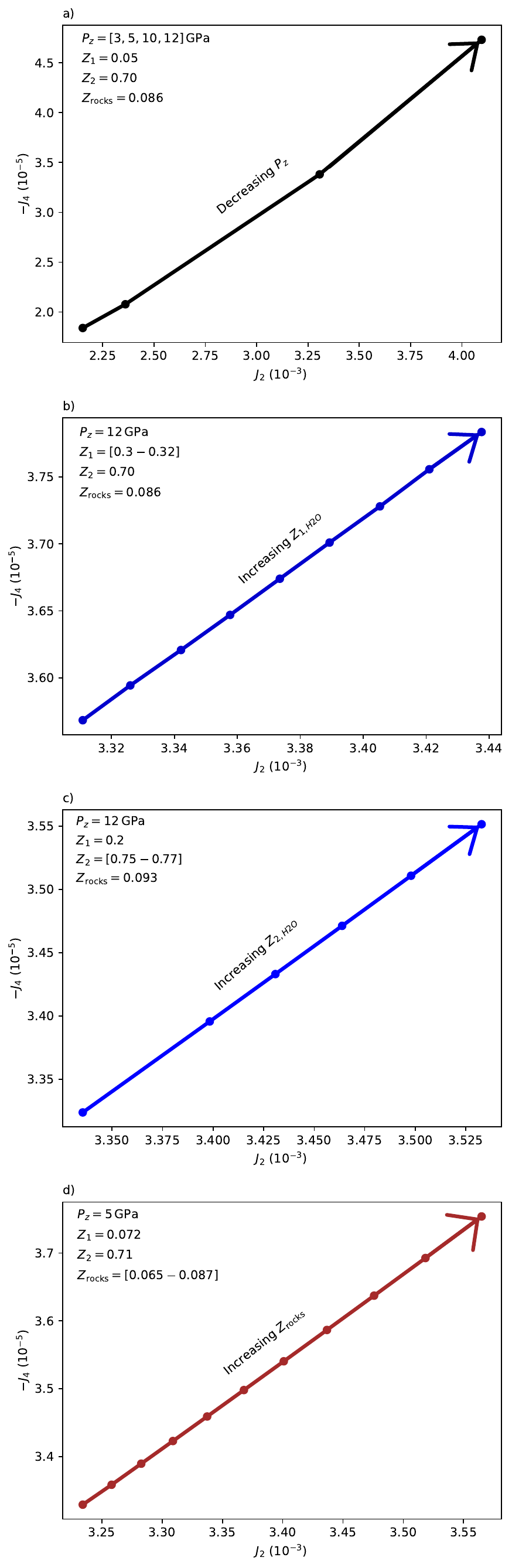}
  \caption{Response of the $J_2$, $J_4$ to variations of a single parameter of case (iv) models. In plot (a) we keep all abundances constant and only vary $P_Z$. In (b) we show the behaviour of the harmonics when varying only the top layer water abundance. In (c) we vary deep water abundance as done in case (ii) models and in (d) we vary rock abundance in the deep interior.}
  \label{fig:freeparametersj2n}
\end{figure}

    We present an additional parameter study to inspect the effects of $P_Z$, as well as the water and rock abundances, on the computed gravity harmonics of Uranus. Figure \ref{fig:freeparametersj2n} shows the dependency of each input parameter whilst the rest remain constant. Decreasing the transition pressure from the water-poor to the water-rich envelope tends to overestimate the harmonics, as shown in Figure \ref{fig:freeparametersj2n}a. This behaviour was seen already in Figure \ref{fig:Js}. When only the envelope water abundance is varied, like in Figure \ref{fig:freeparametersj2n}b or c, this also increases  $J_2$ and $J_4$. Similarly, increasing the amount of rocks in the deeper interior has the same effect.

\end{appendix}
\end{document}